\documentclass{aa}  

\usepackage{graphicx}
\usepackage{txfonts}
\usepackage{hyperref}

\begin{document}

   \title{Imaging the event horizon of M87* from space on different timescales}

   \author{A. Shlentsova\inst{1,2}
          \and
          F. Roelofs\inst{3,4,1}
          \and
          S. Issaoun\inst{3,5,1}
          \and
          J. Davelaar\inst{6,7,1}
          \and
          H. Falcke\inst{1,8,9}
          }

   \institute{Department of Astrophysics, IMAPP, Radboud University, PO Box 9010, 6500 GL Nijmegen, The Netherlands
        \and
             Instituto de Astrof\'isica, Facultad de F\'isica, Pontificia Universidad Cat\'olica de Chile, Casilla 306, Santiago 22, Chile
         \and
             Center for Astrophysics $|$ Harvard \& Smithsonian,  60 Garden Street, Cambridge, MA 02138, USA
        \and
            Black Hole Initiative, Harvard University, 20 Garden Street, Cambridge, MA 02138, USA
        \and 
            NASA Hubble Fellowship Program, Einstein Fellow
         \and
            Center for Computational Astrophysics, Flatiron Institute, 162 Fifth Avenue, New York, NY 10010, USA
         \and
             Department of Astronomy and Columbia Astrophysics Laboratory, Columbia University, 550 W 120th St, New York, NY 10027, USA
         \and
             Max-Planck-Institut für Radioastronomie, Auf dem Hügel 69, D-53121 Bonn, Germany
         \and
             ASTRON, The Netherlands Institute for Radio Astronomy, Postbus 2, NL-7990 AA Dwingeloo, The Netherlands
             }

   \date{Received June 16, 2023; accepted February 4, 2024}

 
  \abstract
   {The concept of a new space very long baseline interferometry (SVLBI) system named the Event Horizon Imager (EHI) has been proposed to dramatically improve black hole imaging and provide precise tests of the theory of general relativity.}
   {This paper presents imaging simulations for the EHI. We investigate the ability to make high-resolution movies of the black hole shadow and jet launching region around the supermassive black hole M87* and other black hole jets with a three-satellite EHI configuration. We aim to identify orbital configurations to optimize the $uv$-coverage to image variable sources.}
   {Observations of general relativistic magnetohydrodynamics (GRMHD) models were simulated for the configuration, consisting of three satellites in circular medium earth orbits with an orbital plane perpendicular to the line of sight. The expected noise was based on preliminary system parameters. Movie frames, for which a part of the $uv$-coverage may be excessively sparse, were reconstructed with algorithms that recover missing information from other frames. Averaging visibilities accumulated over multiple epochs of observations with an appropriate orbital configuration then improves the image quality. With an enhanced signal-to-noise ratio, timescales of observed variability were decreased.}
   {Our simulations show that the EHI with standard system parameters is capable of imaging the variability in the M87* environment on event horizon scales with approximately a month-long temporal resolution. The EHI with more optimistic noise parameters (enhancing the signal-to-noise ratio about 100-fold) would allow for imaging of the variability on gravitational timescales. Observations with an EHI setup at lower frequencies are capable of imaging the variability in extended jets.}
   {Our study shows that the EHI concept can be used to image the variability in a black hole environment and extended jets, allowing for stronger tests of gravity theories and models of black hole accretion, plasma dynamics, and jet launching.}


   \maketitle
%

\section{Introduction} \label{sec:introduction}
    The supermassive black hole in the nucleus of the elliptical galaxy M87 (M87*) has been an object of great interest for intensive studies with imaging observations since its discovery in 1978 \citep{Young1978, Sargent1978}. Measurements of the black hole mass $M_{BH} = (6.5\pm 0.2{| }_{\mathrm{stat}}\pm 0.7{| }_{\mathrm{sys}}) \times 10^9 M_{\odot}$ extracted from the direct measurements of the angular diameter of the shadow \citep{EHT2019a, EHT2019b, EHT2019c, EHT2019d, EHT2019e, EHT2019f} are consistent with the presence of a central Kerr black hole. The estimated mass from the Event Horizon Telescope (EHT) observations agrees with stellar-dynamic observations for which the most recently obtained result is $M_{BH} = (6.6\pm 0.4) \times 10^9 M_{\odot}$ \citep{Gebhardt2011}. \citet{Falcke2000} introduced the black hole `shadow' — the region on the sky plane of a black hole in which there is a noticeable deficit of the observed intensity due to gravitational lensing effects. They proposed that this shadow is resolvable with very long baseline interferometry (VLBI) at sub-millimetre wavelengths. The shadow is caused by two effects: the strong gravitational redshift and a shorter total path length of photons escaping to the observer from geodesics intersecting the horizon, while photons on geodesics missing the horizon can orbit the black hole near the circular radius several times, which leads to a higher integrated emissivity \citep{Bronzwaer2021}. This region is a circle of radius $\sqrt{27}R_g$ (where $R_g = GM_{BH}/c^2$, $G$ is the gravitational constant, $M_{BH}$ is the mass of the black hole and $c$ is the velocity of light) in the Schwarzschild case (the dimensionless black hole spin parameter $a_{*} = 0$) and has a more flattened shape of a similar size for a Kerr black hole. Numerical studies on the observability of the M87* shadow were either done using semi-analytical models \citep{Broderick2009} or with general relativistic magnetohydrodynamics (GRMHD) simulations \citep{Dexter2012, Moscibrodzka2016, Moscibrodzka2017, Ryan2018, Chael2019m, Davelaar2019}. The shadow is directly connected to the time-like component of the space-time metric that is also probed by gravitational wave measurements \citep{Psaltis2020, Psaltis2021}. \\
    \indent M87* has a black hole shadow with an angular size of $\sim42~\mu$as \citep{EHT2019a}. It is the second-largest black hole shadow on the sky as seen from Earth, after the Galactic Centre black hole Sagittarius A* (Sgr\,A*), the supermassive black hole at the centre of the Milky Way. Sgr\,A* has a black hole mass of $M_{BH} = (4.152 \pm 0.014) \times 10^6 {M}_{\odot}$ and an angular shadow size of $\sim52~\mu$as \citep{Doeleman2008, Johnson2015, Lu2018, GRAVITY2019, EHT2022a, EHT2022b, EHT2022c, EHT2022d, EHT2022e, EHT2022f}. M87* has several advantages for observations in comparison with Sgr\,A*, namely the absence of scattering by the interstellar medium between us and the central radio source. The timescales of structural variations of a black hole are characterized by the gravitational time $t_{\mathrm{G}} = GM_{BH}/c^3$ and the innermost stable circular orbital time $t_{\mathrm{ISCO}} \approx 24.25 t_{\mathrm{G}}$ in the case of the rapidly spinning Kerr black hole ($a_{*} = 0.94$). The timescales of structural variations of M87* are, therefore, $\sim10^3$ longer than those of Sgr\,A*, allowing us to assume a static source during single-day observations and make use of classical aperture synthesis techniques. Another striking difference is that M87* hosts a powerful relativistic jet observed in the radio, optical, and X-ray bands \citep[e.g.][]{Sparks1996, Marshall2002}. Although Sgr\,A* may also host a jet, it has not been clearly observed yet. Altogether this makes M87* an auspicious source for observations with VLBI at millimetre and sub-millimetre wavelengths. \\
    \indent Multiple VLBI imaging studies of M87* have been carried out at 3.5~mm (a frequency of 86~GHz), 7~mm (43~GHz), 1.3~cm (24~GHz), and longer wavelengths \citep[e.g.][]{Hada2016, Kravchenko2020, Chang2010}. The development of high-bandwidth VLBI systems made it possible to decrease observational wavelength down to 1.3~mm (230~GHz) for the EHT and reach an instrument angular resolution of about 25~$\mu$as \citep{EHT2019b}. Therefore, the EHT array can resolve the shadow of M87* \citep{EHT2019a, EHT2019b, EHT2019c, EHT2019d, EHT2019e, EHT2019f}. \\
    \indent The EHT data provided the first image of the black hole shadow with the surrounding ring of emission. This allowed for the exclusion of some categories of black hole models \citep{EHT2019e}. With the addition of linear- and circular-polarimetric images of the ring, magnetically arrested disk (MAD) models \citep{Narayan2003} with $a_{*}$ roughly around -0.5 and 0.0 were found to be favoured \citep{EHT2019g, EHT2019h, EHT2019i}. Nevertheless, conclusions are still strongly dependent on the imaging assumptions. A further reduction of acceptable models requires more accurate spin measurements as well as an improved estimation of the magnetization and electron temperature distribution. Quantitative spin measurements require higher-resolution imaging of the source \citep{VanderGucht2019, Roelofs2021}. Measurements of the photon subring parameters of M87* could offer precise measurements of the black hole mass and spin \citep{Johnson2020}. In addition, more accurate measurements of the static photon ring allow for deeper tests of the Kerr metric and consequently the theory of general relativity (GR) \citep{Vagnozzi2022}. Moreover, the plasma behaviour in a black hole environment is currently not fully understood. Imaging the variability of the emitting plasma would provide constraints on the plasma parameters. Hence imaging variability in the M87* environment with a higher resolution and fidelity would deliver data for more accurate reconstructions of the plasma surrounding the black hole, as well as GR tests. Furthermore, such measurements could be used to test different theories of gravity \citep{Mizuno2018} and other non-Kerr alternatives such as boson stars \citep{Olivares2020}, axion models \citep{Chen2020}, or fuzzballs \citep{Bacchini2021}. However, this requires improvements to be made to the EHT. \\
    \indent Earth-based extensions of the EHT array, such as the next-generation EHT \citep[ngEHT,][]{Doeleman2019, Raymond2021, Johnson2023, Doeleman2023}, will substantially improve imaging with increased sensitivity and the additional VLBI baseline coverage from new sites. The ngEHT will also have an expanded frequency coverage up to 345~GHz. Nevertheless, angular resolution improvements for ground-based EHT extensions are limited to a maximum physical baseline length of one Earth diameter. Besides, frequencies higher than 345 GHz are accessible only for very few sites with excellent atmospheric conditions, which makes VLBI image reconstruction at these frequencies difficult, if not impossible. \\
    \indent Further enhancement of the EHT can be achieved with the deployment of EHT antennas into space. The first observations which included space-ground baselines were performed in 1986 by the Tracking and Data Relay Satellite System (TDRSS) and ground-based telescopes in Australia and Japan \citep{Levy1986}. One of the first space VLBI (SVLBI) missions to include space-ground baselines was the VLBI Space Observatory Programme (VSOP) carried by the Highly Advanced Laboratory for Communications and Astronomy (HALCA) satellite that operated in orbit in the period from 1997 to 2003 \citep{Hirabayashi1998, Hirabayashi2000}. Another SVLBI mission of this kind was RadioAstron, which was operational between 2011 and 2019 \citep{Kardashev2013}. \\
    \indent In addition to the EHT array, VLBI stations could be located in a nearly circular Low Earth Orbit \citep{Palumbo2019} or a Highly Elliptical Orbit \citep{Andrianov2021}. Such a setup would provide fast baseline coverage, allowing for dynamical imaging of rapidly varying sources, as Sgr\,A*. VLBI stations located in an elliptical medium Earth orbit (MEO) would increase the angular resolution and the imaging capabilities of the array to distinguish different space-time metrics around Sgr\,A* \citep{Fromm2021}. Inclusion of one telescope in a high-inclination MEO or Geosynchronous Orbit would increase the angular resolution and the number of available sources for which the shadow could be resolved \citep{Fish2019}. For example, the central black holes in IC\,1459 and M84 (NGC\,4374) with predicted shadow diameters of $\sim9.2~\mu$as and $\sim9.1~\mu$as, respectively, as well as the Sombrero Galaxy (NGC\,4594, M104) with an estimated shadow diameter of $\sim5.7~\mu$as and several other sources \citep{Johannsen2012}. The Black Hole Explorer (BHEX\footnote{\url{https://www.blackholeexplorer.org/}}) mission concept aims to launch a telescope into a near-geosynchronous orbit to make precision measurements of a black hole's photon ring properties using ground-space baselines \citep[][Johnson et al., Marrone et al., in prep., SPIE]{Kurczynski2022}, which would complement high-frequency space-space baseline imaging with the Event Horizon Imager concept presented in this work. A station on the Moon or in the Earth-Sun or Earth-Moon Lagrange points, added to the EHT array, would sharpen the angular resolution sufficiently to resolve signatures of the substructures in the photon rings of M87* and Sgr\,A* \citep{Johnson2020}. It was proposed that the Origins Space Telescope, equipped with a modified version of its heterodyne instrument, could be used for these purposes \citep{Pesce2019}. \\
    \indent An alternative approach is to consider only space-space baselines. Such a system has the advantage that atmospheric data corruptions can be avoided, thus allowing for observation at higher frequencies in addition to the possibility of reaching baselines longer than an Earth diameter. Therefore, the resolution can be increased even further. \\
    \indent The Event Horizon Imager (EHI) is a concept for a new SVLBI system that consists of two or three satellites in polar or equatorial circular MEOs observing with only space-space baselines at frequencies up to 690~GHz \citep{Martin-Neira2017, Kudriashov2019, Roelofs2019, Kudriashov2021b}. The concept envisages the precise localization of the interferometer baseline based on the real-time relative positioning of the satellites using the Global Navigation Satellite System (GNSS). In addition, the concept suggests the on-board cross-correlation for the reduction of the difficulties of the data downlink to the ground. The accurate positioning of the satellites and interchange of local oscillator signals permit the phase calibration of EHI observations and, thus, the use of complex visibilities \citep{Kudriashov2021b}. Both the proposed on-the-fly positioning and the on-board data processing require a working inter-satellite link (ISL) to perform highly accurate ranging measurements, as well as to exchange observed signals and local oscillator components for coherent operation \citep{Martin-Neira2019, Kudriashov2021a}. This implies a direct line of sight between the satellites. Hence the maximum baseline length, limiting the angular resolution of the EHI, is set by the radius of the satellite orbits and the occultation of the ISL by the Earth. The proposed resolution is $\sim5$~$\mu$as \citep{Kudriashov2021b}, which is an order of magnitude better than what can be obtained with the EHT from Earth. Due to higher observation frequencies, the EHI can detect emission that originates from regions that are closer to the black hole \citep{Moscibrodzka2009}. Those regions are more dominated by general relativistic effects, leading to the reduced variability of the images. Additionally, it causes the emission to trace the photon ring more closely. Moreover, the proposed system can allow one to avoid a significant part of the interstellar scattering for the observations of Sgr\,A* since the scattering kernel size decreases with the square of the observing frequency \citep{Roelofs2019}. \\
    \indent Simulated observations of Sgr\,A* with the EHI demonstrate the excellent imaging capability with this concept \citep{Roelofs2019}. These simulations have shown that the EHI could image Sgr\,A* with an order of magnitude higher angular resolution and accuracy than the EHT within just a few months of observation, assuming standard system noise parameters. \\
    \indent The overarching EHI science goal is to test theories of gravity. Precise imaging of a black hole photon ring brings essential information to distinguish between GR and alternative theories. The imaging of the plasma variability would provide constraints on models of plasma dynamics and jet formation. Therefore, the motivation for further testing of the EHI capabilities is to study aspects limiting the fidelity of imaging on different timescales of observations in order to determine how accurately GR, as well as models of black hole accretion, plasma dynamics and jet launching, can be tested with the EHI. In addition, understanding EHI imaging constraints can help to optimize the design of the system. \\
    \indent This paper supplements previous EHI studies by examining several configurations of the system. We study their influence on the possibility of resolving structural variations of the M87* environment and extended jets of other black holes (e.g. NGC\,1052) at millimetre and sub-millimetre wavelengths. The research is focused on M87* since its timescales of structural variations are significantly longer than those of Sgr\,A*, as mentioned above. The considered system setups are described in Section~\ref{sec:setup}. Section~\ref{sec:generation} introduces source models and image generation. The simulation results are presented in Section~\ref{sec:results}, and conclusions are summarized in Section~\ref{sec:conclusions}.

\begin{figure}
\resizebox{\hsize}{!}{\includegraphics{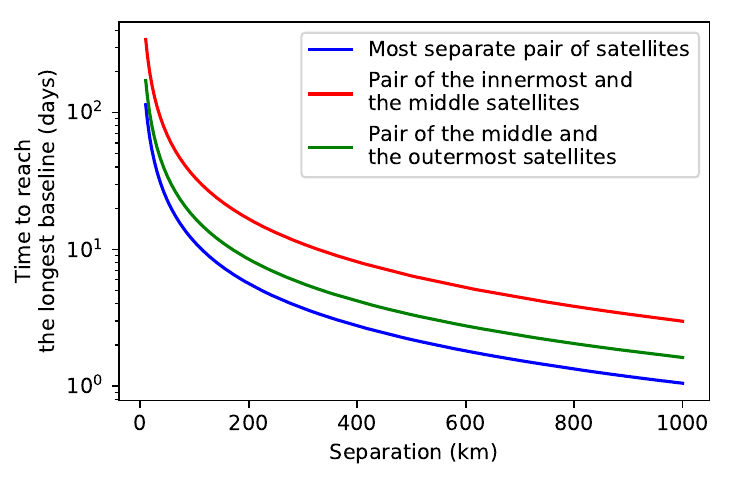}}
\caption{Time to reach the longest baseline as a function of orbital separation for the considered EHI system setup. Blue line corresponds to the most separate pair of satellites; red line corresponds to the pair of the innermost and the middle satellites; green line corresponds to the pair of the middle and the outermost satellites.}
\label{fig:septime}
\end{figure}

\begin{figure*}
\centering
\includegraphics[width=17cm]{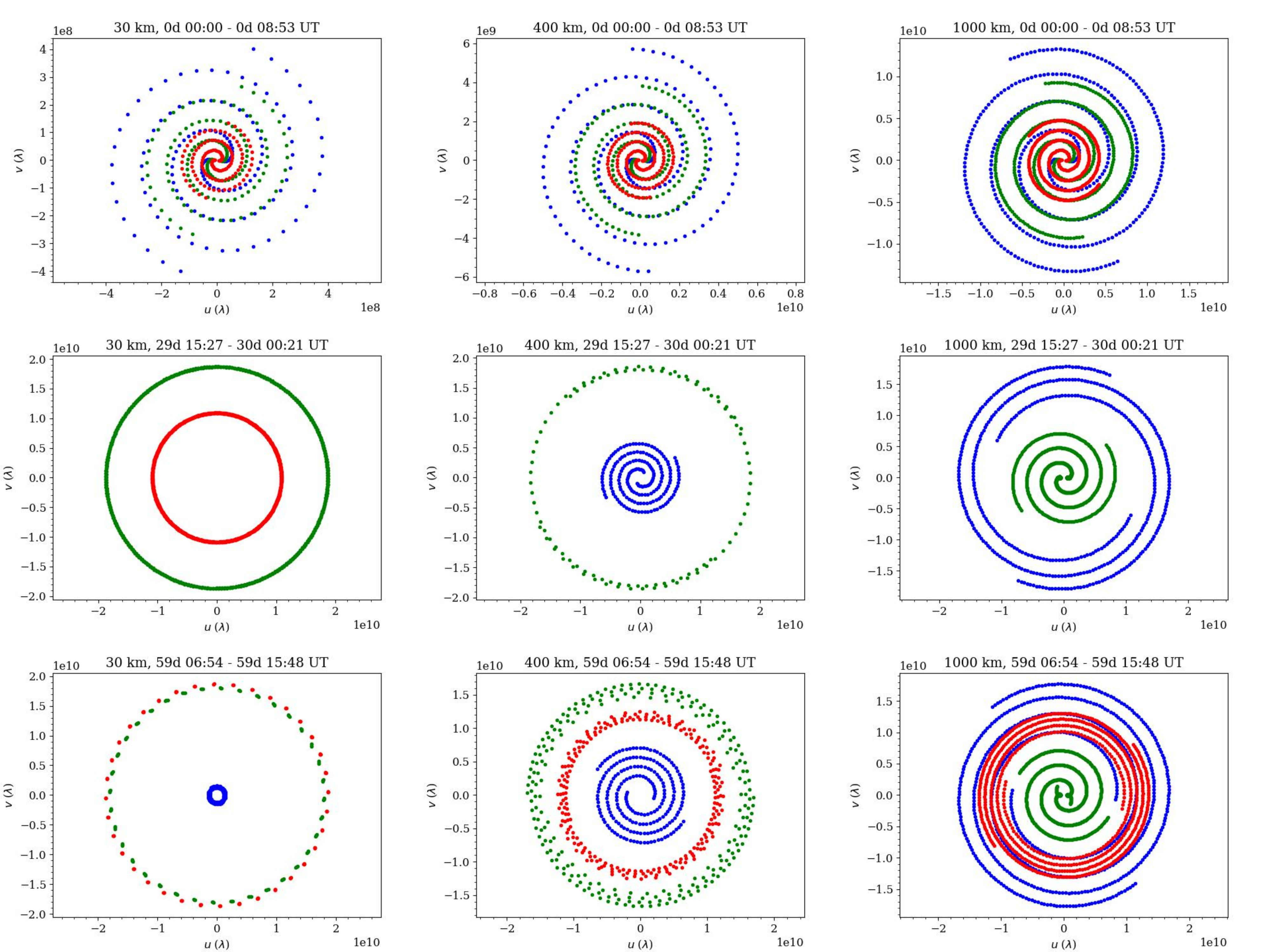}
\caption{Parts of the $uv$-coverage of 8.9-hour duration for 30, 400, and 1000~km separations of the orbits at 230~GHz; first, 81st, and 161st snapshots are shown (from top to bottom). Blue points correspond to the most separate pair of satellites; red points correspond to the pair of the innermost and the middle satellites; green points correspond to the pair of the middle and the outermost satellites.}
\label{fig:uvsnap}
\end{figure*}

\section{EHI system setup} \label{sec:setup}
    The EHI concept implies two or three satellites in polar or equatorial circular MEOs, as mentioned above. The particular configuration is still under discussion. The EHI system considered in this paper consisted of three satellites in circular MEOs with slightly different radii. The largest radius of the satellite orbits was 13\,913~km. The selected radius provides the longest possible baselines taking into account the required simultaneous accessibility of at least three GNSS satellites from the EHI orbits. At the same time, this maximal radius ensures the orbits of all three EHI satellites to be between the Van Allen belts \citep{Kudriashov2019}. The second satellite was placed on an orbit with a radius smaller by a distance referred to hereafter as the orbital separation. The third satellite was added between the first two. Its orbit radius was one-third of the orbital separation bigger than for the inner satellite. The orbital plane was set perpendicular to the line of sight to the observed source, taking into account the declination. In reality, this condition without additional reorientation of the orbital plane can be fulfilled only for a very limited list of sources. Due to the different orbital radii and velocities of the satellites, the complete $uv$-coverage will have the shape of a spiral. We investigated systems with constant orbital separations of 30, 50, 60, 100, 200, 300, 350, 400, 500, and 1000~km. \\
    \indent The imaging of the horizon-scale variability requires satisfactory coverage of the whole $uv$-plane over time intervals comparable to the timescale of the investigated variability. The longest available baseline is limited by the radius of the satellite orbits and the occultation of the ISL by the Earth. Let us assume that all three satellites start at the same orbital phase. When the first pair of satellites reaches the longest baseline, the other two pairs are also at the relatively long baselines. If the system continues observation without configuration changes, the $uv$-coverage provided by the remaining pairs of satellites lacks short baselines. This situation continues until the occultation of the ISL between the first pair of satellites is finished and they converge again. During this time, the other two pairs of satellites sequentially find themselves in a similar situation, which creates a prolonged period of increased $uv$-coverage sparseness. This sparsity period can in principle be avoided by interchanging the satellite orbits each time when the longest baseline length is reached. Figure~\ref{fig:septime} shows that the time to reach the longest baseline depends on the orbital separation and rapidly decreases from approximately a month to several days. Regular changes in the satellite orbits on such short timescales are exceedingly fuel-consuming and, therefore, unrealistic. The exclusion of satellite orbit changes will save fuel, resulting in a prolongation of the overall mission duration and an increase in both the amount of collected data and the number of objects that can be observed. We considered satellites to stay in their orbits as long as required to get a long enough series of $uv$-data points. The full data series can be used for the reconstruction of a highly detailed time-averaged image. In this paper, we divided the series of data into parts to obtain the time resolution of interest. Figure~\ref{fig:uvsnap} illustrates this concept with an example (see Sec.~\ref{sec:uv} for details). Therefore, the temporal resolution did not depend on orbital separation and was selected during data processing. \\
    \indent In this work, we considered the standard EHI system composed of 4.0-metre antennas, fitting in the Ariane 6 spacecraft \citep{Arianespace2016}, and a hypothetical EHI+ system consisting of 15.0-metre antennas \citep[see also][]{Gurvits2022}. System parameters for the noise addition were taken from \citet{Roelofs2019} for the EHI system. The EHI+ system was assumed to have a system temperature close to the minimum allowable for a coherent heterodyne system and a wider observing bandwidth to increase the signal-to-noise ratio (Sec.~\ref{sec:noise}). In addition, EHI+ antennas were considered to observe at higher frequencies (see Sec.~\ref{sec:maps}). Technical specifications assumed in this paper for EHI+ are foreseen by the authors as technically very challenging and not yet demonstrated, but not impossible. Phase corruptions due to uncertainties in the orbital model have not been studied in this work; it was assumed that the baseline vector is known exactly thanks to the GNSS-assisted orbit determination, inter-satellite link ranging measurements, a local oscillator sharing setup, and other measures that are currently under investigation \citep{Kudriashov2021b}. Image reconstructions were therefore produced using complex visibilities; the extent of the necessity of using closure products will depend on the eventual feasibility and performance of the connected-interferometer concept laid out in \citet{Kudriashov2021b} (see also Sec. \ref{sec:uv}). An investigation of the influence of the detailed system noise and calibration parameters on the reconstructed images presented here is left for future work.

\section{Simulated observations} \label{sec:generation}
    In this section, the generation of simulated observations is described. Theoretical emission maps were produced from GRMHD simulations and used as input. These source models were used to calculate complex visibilities for the simulated observations together with the coverage of the $uv$-plane, produced by the system setup outlined in Section~\ref{sec:setup}. The calculated complex visibilities were supplemented by data corruption effects (e.g. thermal noise) and then reconstructed with a Regularized maximum likelihood (RML) algorithm. The quality of the image reconstructions was quantified by several image quality metrics.

\begin{table*}
\caption{Parameters of the theoretical emission maps used.}
\label{tab:models}
\centering
\begin{tabular}{c c c c c c}
\hline\hline
source & time interval (hours) & movie duration (months) & $\nu$ (GHz) & total flux (Jy) & FoV ($\mu$as) \\
\hline
&&& 230 & 0.84 - 1.31 & 190.9 \\
M87* & 89 (10~$t_{\mathrm{G}}$) & 10 (800~$t_{\mathrm{G}}$) & 560 & 0.47 - 0.84 & 190.9 \\
&&& 5000 & 0.10 - 0.18 & 190.9 \\
&&& 43 & 0.60 - 0.82 & 763.7 \\
\hline
NGC\,1052 & 89 & 10 & 43 & 7.0 - 7.6 & $9.6 \times 10^3$ \\
\hline
&&& 230 & 0.93 - 1.04 & 190.9 \\
M87* & 8.9 (1~$t_{\mathrm{G}}$) & 2.5 (200~$t_{\mathrm{G}}$) & 560 &  0.48 - 0.56 & 190.9 \\
&&& 5000 & 0.10 - 0.12 & 190.9 \\
&&& 43 & 0.69 - 0.72 & 763.7 \\
\hline
NGC\,1052 & 8.9 & 2.5 & 43 & 7.39 - 7.46 & $9.6 \times 10^3$ \\
\hline
\end{tabular}
\tablefoot{Duration of movies was calculated as time intervals between frames multiplied by the number of frames in sets. For $uv$-coverage calculation, the same time intervals between frames in movies were used as for M87* since the black hole mass was not properly scaled for NGC\,1052 models when changing the angular size of the source.}
\end{table*}

\subsection{Theoretical emission maps} \label{sec:maps}
    Modelling of accretion flows around black holes is typically done by performing GRMHD simulations \citep[see][and references therein]{Porth2019}. The majority of these simulations only solve for the dynamically important protons, although more recent works also include information on the electron population via either including electron thermodynamics in GRMHD \citep{Ryan2018, Chael2019m} or by using Particle-in-Cell simulations instead \citep{Parfrey2019, Crinquand2021, Bransgrove2021}. The GRMHD models used in this work do not contain information on the electron population, and therefore use a parametrization for their properties, such as the shape of the distribution function, in the subsequent ray tracing. In this work, we used the $\kappa$-jet model of M87* from \citet{Davelaar2019}, first developed for Sgr\,A* in \citet{Davelaar2018}. The $\kappa$-jet model for M87* is capable of recovering the observed spectral energy distribution from radio to near-infrared, the jet core shift relation and shows consistent image morphologies at 230~GHz. The model assumes that the electron distribution function is a $\kappa$-distribution function, which is a combination of a thermal core and a power-law. The slope of the power-law is set by a parametrization from \citet{Ball2018}, who studied trans-relativistic reconnection. The parameterization effectively adds a power-law population in the jet sheath, while electrons in the disk are in a thermal distribution. \\
    \indent The dynamics of the accretion flow onto the black hole were simulated using the Black Hole Accretion Code \citep[BHAC,][]{Porth2017, Olivares2019} that solves the GRMHD equations. The performed simulation assumed a standard and normal evolution (SANE) accretion disk, the dimensionless black hole spin parameter was set to be $a_{*} = 15/16$. The general relativistic ray-tracing code RAPTOR \citep{Bronzwaer2018, Bronzwaer2020} was used to generate synthetic synchrotron maps \citep[for further details, see][]{Davelaar2019}. The resulting collections of synthetic synchrotron maps of the jet-launching region in M87* were used in this paper as input models to simulate observations. These maps were calculated at an inclination of $i = 160^{\circ}$, which ensured that the orientation of the emitting region corresponds to the results of the EHT observations \citep{EHT2019a, EHT2019e}. The collection of synthetic synchrotron maps represents a black hole environment with a certain time interval between frames, defined by the gravitational timescale of a black hole $t_{\mathrm{G}}$. In this work, the models had 1 and 10~$t_{\mathrm{G}}$ (8.9 and 89 hours, respectively) intervals between frames. \\
    \indent The analysis for the EHI system was carried out for three frequencies, namely 43~GHz, which is a standard frequency for many VLBI observations, 230~GHz, which is the operating frequency of the EHT, and 560~GHz, which is an expected operating frequency of the EHI. 560~GHz was selected, following one of the secondary EHI science goals of imaging the water line in protoplanetary disks. If future technical studies find that ground telescope support will be a system requirement, this observing frequency may be adjusted to, for example, 690 GHz without large consequences for the observable black hole shadow features. \\
    \indent In the case of the EHI+ system, the analysis was performed for two frequencies, which are 560~GHz and 5~THz, to test the imaging limitations of the EHI concept. The frequency of a few THz was selected since this additionally increases the angular resolution and, therefore, the number of sources for which a black hole shadow is resolvable, for instance, one in Centaurus~A \citep{Janssen2021}. It should be noted that 5~THz corresponds to a wavelength of 60~microns, which translates to very stringent requirements for the phase calibration and antenna surface accuracies of the EHI+ system. To get models at 5~THz, synthetic synchrotron maps initially calculated at 560~GHz were scaled to the flux at 5~THz according to the flux density spectrum. The flux density spectrum in the $\kappa$-jet model is described by a power-law $F_{\nu}\propto\nu^{\alpha}$ with index $\alpha\approx-0.7$ at high frequencies ($\nu>230$~GHz), as declared in \citet{Davelaar2019}. \\
    \indent Another potential secondary EHI science goal is imaging the structural variations of the extended jet of black holes at 43~GHz. Many active galactic nuclei (AGNs) are located further away or have smaller black hole masses than M87*, thus, shadows of these black holes can not be resolved even with the EHI. Nevertheless, they are suitable for imaging relativistic jets on larger scales. The considered M87* model provides information about the jet on scales much shorter than in the observations. Hence additional synthetic synchrotron maps with $i = 90^{\circ}$ inclination of the source were scaled so that the field of view and the total flux correspond to the parameters of the jet of another AGN. We note that this scaling is not physical because the black hole mass and hence the variability timescales are not properly scaled when changing the angular size of the source. Nevertheless, these coarse maps allow one to get insight into the ability to resolve AGN jet structures with the EHI. As a generic example for an AGN jet, NGC\,1052 was chosen \citep{Kadler2004, Ros2008, Baczko2019, Nakahara2019, Gurvits2019}. Table~\ref{tab:models} summarizes parameters of collections of synthetic synchrotron maps that are used in this work.

\subsection{Coverage of the $(u, v)$ plane} \label{sec:uv}
    Calculation of the complex visibilities was performed with the \verb|eht-imaging| software \citep{Chael2018, Chael2019}. This requires, besides the source model, a $uv$-coverage as input. As described in Section~\ref{sec:setup}, we fixed the satellite orbits throughout the observations. The obtained $uv$-data was split in the time domain into snapshots of a duration corresponding to the time interval between frames in the theoretical emission maps, namely 1 and 10~$t_{\mathrm{G}}$ (8.9 and 89 hours). Therefore, each part of the $uv$-coverage was related to the specific frame of the source model when calculating the complex visibilities. \\
    \indent Following \citet{Roelofs2019}, we set an integration time per measurement that is $uv$-distance-dependent,
    \begin{equation}\label{eq:tint}
        t_{\mathrm{int}} = \frac{P}{4 \pi D_{\lambda} \Theta},
    \end{equation}
    where $P$ is the orbital period of the innermost satellite, $D_{\lambda}$ is the length of the corresponding baseline, and $\Theta$ is the field of view (FoV) of the EHI system. Thereby $t_{\mathrm{int}}$ is within the $uv$-smearing limit, so we can avoid the corruption of the reconstructed image due to the displacement of the $uv$-vector during an integration time \citep{Thompson2017, Palumbo2019}. For the shortest baselines, the integration time was chosen so that the $uv$-arcs are limited to 10 degrees. The FoV of the EHI system was chosen corresponding to the FoV, provided by the model. \\
    \indent The time it takes to reach the longest baselines (Figure~\ref{fig:septime}) and the spiral density depend on the orbital separation. Hence the described $uv$-coverage after splitting into snapshots demonstrates three distinctive possibilities of points distribution on the $uv$-plane for each pair of satellites. Figure~\ref{fig:uvsnap} illustrates these possibilities with an example of several snapshots for three different orbital separations. The first option is a `spiral-like' distribution. For this option, it is typical that the points are allocated sparsely but homogeneously in the shape of a part of a spiral (e.g. Figure~\ref{fig:uvsnap}, middle right panel). The second option is a `ring-shaped' distribution. The points are allocated tightly to each other, however, they are grouped into a narrow ring (e.g. Figure~\ref{fig:uvsnap}, middle left panel). The third option is an intermediate one between the previous two. For this option, it is typical that a part of a spiral forms a ring of a medium width (Figure~\ref{fig:uvsnap}, bottom middle panel). With a small orbital separation, a spiral-like distribution is present only in the first snapshot (Figure~\ref{fig:uvsnap}, top left panel). In the other snapshots, all three pairs of satellites demonstrate a ring-shaped distribution of points on the $uv$-plane (Figure~\ref{fig:uvsnap}, middle left and bottom left panels). With an increasing orbital separation, the $uv$-coverage from each pair of satellites expands through an intermediate phase to a spiral-like distribution (Figure~\ref{fig:uvsnap}, bottom panels). \\
    \indent The coverage of the $uv$-plane formed by the whole system is one, two or three concentric narrow rings, depending on how many pairs of satellites are blocked by the Earth, in most of the snapshots for the small orbital separations. Therefore, the resulting $uv$-coverage provides information only on a very limited range of baselines in each snapshot. When the orbital separation increases, the range of baselines represented in each snapshot also grows due to wider rings, providing more homogeneous coverage. As a trade-off, the shortest baselines become longer and the coverage becomes less dense. Hence optimizing the movie reconstruction quality requires finding a balance between density and isotropy of the $uv$-coverage in each snapshot, taking the source evolution timescales into account. Nevertheless, with a suitable orbital separation, sufficient coverage of the $uv$-plane can be obtained without additional changes in the satellite orbits during the observations. \\
    \indent Similar to the time it takes to reach the longest baselines, the time it takes until the temporally blocked satellites catch up with the other satellites, restoring the ISL as its path is no longer occulted by Earth, is inversely proportional to the orbital separation. Therefore, over multiple iterations, an increase of the orbital separation leads to an increase of the fractional time period with the co-visibility of three satellites, which is necessary for the calculation of closure phases (i.e.~the phase of the bispectrum, the sum of visibility phases on a closed triangle of baselines). For example, with an orbital separation of 50~km, only $35\%$ frames of the 10~$t_{\mathrm{G}}$ duration are suitable for reconstructions with amplitudes and closure phases instead of complex visibilities on the investigated observational period. 100 and 200~km separation provides $45\%$ appropriate frames, while 300 and 400~km orbital separations grant $75\%$ and $90\%$ suitable frames, respectively. Since closure phases are immune to station-based phase errors, and the bispectrum imposes less strict requirements on the phase stability of the interferometer, the use of these robust quantities relaxes technical system requirements for the EHI. Therefore, the dependence of the possibility to calculate closure phases on the orbital separations should be considered in the design of the system.

\begin{table*}
\caption{System parameters and resulting noise.}
\label{tab:noise}
\centering
\begin{tabular}{l|cccc|cc}
\hline\hline
& \multicolumn{4}{c|}{EHI system} & \multicolumn{2}{c}{EHI+ system} \\
& \multicolumn{3}{c}{M87*} & NGC\,1052 & \multicolumn{2}{c}{M87*}\\
\hline
$\nu$ (GHz) & 230 & 560 & 43 & 43 & 560 & 5000 \\
$D$ (m) & \multicolumn{4}{c|}{4.0} & \multicolumn{2}{c}{15.0} \\
\hline
$T_{\mathrm{sys}}$ (K) & \multicolumn{4}{c|}{150} & 50 & 300 \\
$\eta_{\mathrm{ap}}$ & \multicolumn{4}{c|}{0.58} & \multicolumn{2}{c}{0.58} \\
$\eta_{\mathrm{cor}}$ & \multicolumn{4}{c|}{0.97} & \multicolumn{2}{c}{0.97} \\
$\eta_{\mathrm{clock}}$ & \multicolumn{4}{c|}{0.87} & \multicolumn{2}{c}{0.87} \\
$\Delta\nu$ (GHz) & \multicolumn{4}{c|}{5} & 25 & 450 \\
$t_{\mathrm{int,centre}}$ (s) & \multicolumn{4}{c|}{452.2} & \multicolumn{2}{c}{452.2} \\
$t_{\mathrm{int,edge}}$ (s) & 65.67 & 27.12 & 87.82 & 6.99 & 27.12 & 3.02 \\
\hline
$\mathrm{SEFD}$ (Jy) & \multicolumn{4}{c|}{$6.7 \times 10^4$} & \multicolumn{2}{c}{$1.6 \times 10^3$} \\
$\sigma_{\mathrm{centre}}$ (Jy) & \multicolumn{4}{c|}{0.036} & \multicolumn{2}{c}{0.00038} \\
$\sigma_{\mathrm{edge}}$ (Jy) & 0.094 & 0.15 & 0.082 & 0.29 & 0.0016 & 0.0047 \\
\hline
\end{tabular}
\tablefoot{$\sigma$-values are calculated with equations~\ref{eq:noise} and~\ref{eq:sefd} for 30~km orbital separation at the centre (long integration time) and edge (short integration time) of the $uv$-spiral. Integration time at the edge of the $uv$-spiral $t_{\mathrm{int,edge}}$ at 43~GHz is different for NGC\,1052 and M87* due to different fields of view (FoVs).}
\end{table*}

\subsection{System noise calculation} \label{sec:noise}
    Thermal noise in the system can be described as a Gaussian noise in the complex visibility plane with zero mean and standard deviation $\sigma$. In radio interferometry, $\sigma$ can be calculated from the integration time $t_{\mathrm{int}}$, the observing bandwidth $\Delta\nu$ and System Equivalent Flux Densities of the antennas $\mathrm{SEFD}_{1,2}$ \citep{Thompson2017}. For the two-bit sampling of the signal with four-level quantization,
    \begin{equation}\label{eq:noise}
        \sigma = \frac{1}{0.88} \sqrt{\frac{\mathrm{SEFD}_{1} \mathrm{SEFD}_{2}}{2 \Delta\nu t_{\mathrm{int}}}}.
    \end{equation}
    SEFD is standardly defined as
    \begin{equation}\label{eq:sefd}
        \mathrm{SEFD} = \frac{2 k_{\mathrm{B}} T_{\mathrm{sys}}}{\eta A},
    \end{equation}
    where $k_{\mathrm{B}}$ is Boltzmann’s constant, $T_{\mathrm{sys}}$ is the system temperature, $\eta$ is the efficiency and $A = \pi (D/2)^2$ is the area of an antenna with diameter D. The efficiency $\eta = \eta_{\mathrm{ap}} \eta_{\mathrm{cor}} \eta_{\mathrm{clock}}$ includes efficiencies of the aperture, correlator, and clock, respectively. All antennas were assumed to have constant efficiency independently of observational frequency. In practice, the considered efficiency can be implemented for only one frequency, while at higher frequencies the efficiency will be lower. The bandwidth of the ISL for all EHI configurations and EHI+, observing at 560~GHz, was the sum of bandwidths in two polarizations, hence equals $2\Delta\nu$. For EHI+, observing at 5~THz, the bandwidth was the sum of bandwidths in two polarizations and two sidebands, so it equals $4\Delta\nu$. The parameters and resulting noise for the considered system setups are shown in Table~\ref{tab:noise}. The noise increases towards the edge of the spiral due to the $uv$-distance-dependent integration times (see Sec.~\ref{sec:uv}).

\subsection{Image reconstruction} \label{sec:reconstruction}
    With a finite $uv$-sampling, image reconstruction requires imposing assumptions due to the non-uniqueness of images reproducing the observed data. The CLEAN algorithm \citep{Hogbom1974} represents the sky image in terms of a finite number of point sources to constrain the image. Regularized maximum likelihood (RML) algorithms \citep{Gull1978} minimize the weighted sum of the goodness-of-fit test statistic $\chi^2$ and a set of regularizers that favour certain image characteristics such as smoothness, sparsity, or similarity to a prior \citep[e.g.][]{Chael2016, Chael2018}. In this work, two methods of applying the RML algorithm were used: snapshot imaging and dynamical imaging, following \citet{Johnson2017}. \\
    \indent Imaging of the source dynamics implies long-lasting observations, which can be used later for the reconstruction of a movie. This movie should display changes in the surrounding environment of a black hole. In snapshot imaging, a set of images, making up the movie, is reconstructed from a corresponding set of observations, and each reconstruction is performed independently. We selected a combination of the Gull-Skilling entropy function \citep{GullSkilling1991} and the Total Squared Variation regularizer \citep{Kuramochi2018} from several regularization functions that are implemented into the RML algorithm. The selection of parameters for snapshot imaging is explained in Appendix~\ref{sec:snapshot}. \\
    \indent Dynamical imaging assumes that the images in the set are temporally connected, each being a perturbation of the previous frame. This allows one to get information that is lacking for high-quality reconstruction from previous and subsequent frames. In dynamical imaging, we used a combination of the Total Squared Variation regularizer, the Second Moment regularizer \citep{Issaoun2019} and a generic regularizer $\mathcal{R}_{\Delta t}$, enforcing continuity from frame to frame in the reconstructed movie \citep{Johnson2017}. The selection of parameters for dynamical imaging is explained in Appendix~\ref{sec:dynamical}. \\
    \indent We reconstructed movies with three temporal resolutions: 1, 10, and 100~$t_{\mathrm{G}}$ (8.9~hours, 3.7~days, and 37~days). The complex visibilities were calculated from the source model frames and corresponding parts of the $uv$-coverage frame by frame. Then data files with complex visibilities were united according to the assumed temporal resolution of the reconstructed movie. Movies with 10~$t_{\mathrm{G}}$ and 100~$t_{\mathrm{G}}$ temporal resolutions were simulated based on source models with 1~$t_{\mathrm{G}}$ and 10~$t_{\mathrm{G}}$ time intervals between frames, respectively. Thus, the variability of the source throughout the observation was included in the reconstructed movies. For the comparison with the theoretical emission maps, the latter was averaged over the corresponding number of frames.

\subsection{Image quality metrics} \label{sec:qualitymetric}
    Quantitative comparison of the reconstructed image with the true model image was performed via two metrics, namely the Normalized Root-Mean-Square Error (NRMSE) and the Normalized cross-Correlation (NXCORR). The NRMSE evaluates images based on pixel-to-pixel similarities and is given by
    \begin{equation} \label{eq:nrmse}
        \mathrm{NRMSE} = \sqrt{\frac{\sum_{i=1}^{n^2}(I_i - I_i')^2}{\sum_{i=1}^{n^2}I_i^2}},
    \end{equation}
    where $I_i$ is the intensity of the $i$th pixel of the $n \times n$ pixels model image and $I_i'$ is that of the reconstructed image \citep{Chael2016, Chael2018}. For the best fit, the NRMSE should be minimized since it is zero when the reconstructed image is identical to the true image. \\
    \indent The NXCORR (a sliding inner-product of two normalized functions) is determined as the cross-correlation of the Fourier transforms of the normalized intensity patterns of two images at different relative shifts \citep[e.g.][]{EHT2019a, Issaoun2019}. Since the NXCORR compares the bulks of each intensity pattern, it is less sensitive to individual features in the reconstructed image. For a given relative shift $\boldsymbol{\delta}$,
    \begin{equation}\label{eq:nxcorr}
        \mathrm{NXCORR}(\boldsymbol{\delta}) = \mid\mathcal{F}^{-1}\{\mathcal{F}\{I_{\mathrm{norm}}^{*}(\boldsymbol{x})\}\boldsymbol{\cdot}\mathcal{F}\{I_{\mathrm{norm}}'(\boldsymbol{x+\delta})\}\}\mid,
    \end{equation}
    where $\mathcal{F}$ is the Fourier transform operator, $I_{\mathrm{norm}}$ is the normalized intensity pattern of the true image and $I_{\mathrm{norm}}'$ is the same for the reconstructed image. The normalized intensity for each pixel $i$ in the image can be calculated as
    \begin{equation}\label{eq:normint}
    I_{\mathrm{norm},i} = \frac{I_i-\mu_{\mathrm{I}}}{\sigma_{\mathrm{I}}},
    \end{equation}
    where $\mu_{\mathrm{I}}$ and $\sigma_{\mathrm{I}}$ are the mean and standard deviation of the intensity distribution in the image. One of all the different shifts across the extent of the images resulting in the maximum cross-correlation is then used to output the final NXCORR value for the two images. Thus, the maximized NXCORR gives the best fit. \\
    \indent Despite the differences in comparison methods, the NRMSE and the NXCORR demonstrate the same qualitative dependencies (see Appendix~\ref{sec:appendix1}). Therefore, for comparison of movies obtained under different conditions, only NXCORR is shown. To evaluate the quality of a set of images making up the movie, metrics were calculated for each frame independently and then averaged over the duration of the movie.

\section{Simulation results} \label{sec:results}
    In this section, we describe the outcome of the simulations for which the setup is described in the previous sections. We start with observations of the M87* shadow with the EHI at 230 and 560~GHz. Further, we describe observations with the EHI+ at 560~GHz and 5~THz. Finally, observations of AGN jets at 43~GHz are described.

\subsection{Imaging of the M87* shadow} \label{sec:shadow}
    Simulations of M87* shadow observations were produced from models with 1 and 10~$t_{\mathrm{G}}$ time intervals between frames and then reconstructed with temporal resolutions of 10 and 100~$t_{\mathrm{G}}$ respectively. These simulations were reconstructed using both snapshot and dynamical imaging methods. Additionally, observations with 1~$t_{\mathrm{G}}$ temporal resolution were reconstructed using dynamical imaging, except the observations at 5~THz due to the heaviness of the computations at this frequency. In this subsection, we compare the quality of movie reconstructions and investigate the optimal orbital separation.

\begin{figure}
\resizebox{\hsize}{!}{\includegraphics{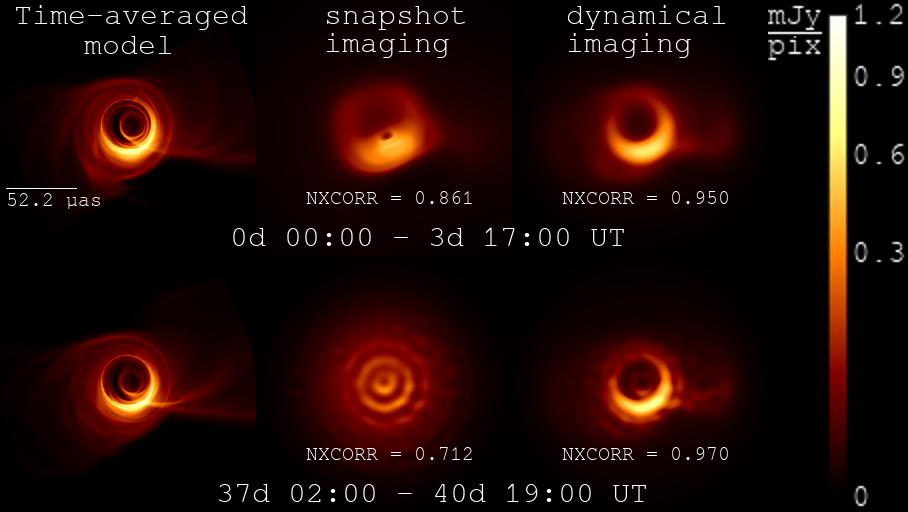}}
\caption{Effect of the reconstruction method on the reconstruction quality. From left to right: the time-averaged over 3.7 days theoretical emission map of M87* at 230~GHz (the model with 8.9~hours between frames); frames of the movies simulated for the EHI with 30~km orbital separation, each lasting 3.7~days, reconstructed independently (snapshot imaging method) and using dynamical imaging \citep{Johnson2017}. The source is varying during the simulated observation. Colours indicate brightness/pixel in mJy (square root scale).}
\label{fig:imgEHImethod}
\end{figure}

\begin{figure}
\resizebox{\hsize}{!}{\includegraphics{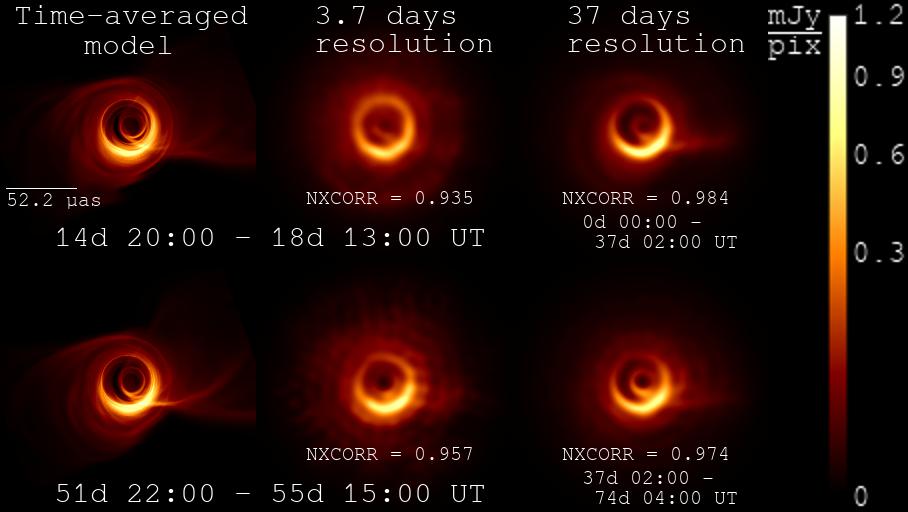}}
\caption{Effect of the temporal resolution on the reconstruction quality. From left to right: the time-averaged over 3.7 days theoretical emission map of M87* at 230~GHz (the model with 8.9~hours between frames); frames of the movies simulated for the EHI with 30~km orbital separation, each lasting 3.7 and 37~days, reconstructed using snapshot imaging (to highlight the effect). The source is varying during the simulated observation. Colours indicate brightness/pixel in mJy (square root scale).}
\label{fig:imgEHItscale}
\end{figure}

\begin{figure*}
\sidecaption
\includegraphics[width=12cm]{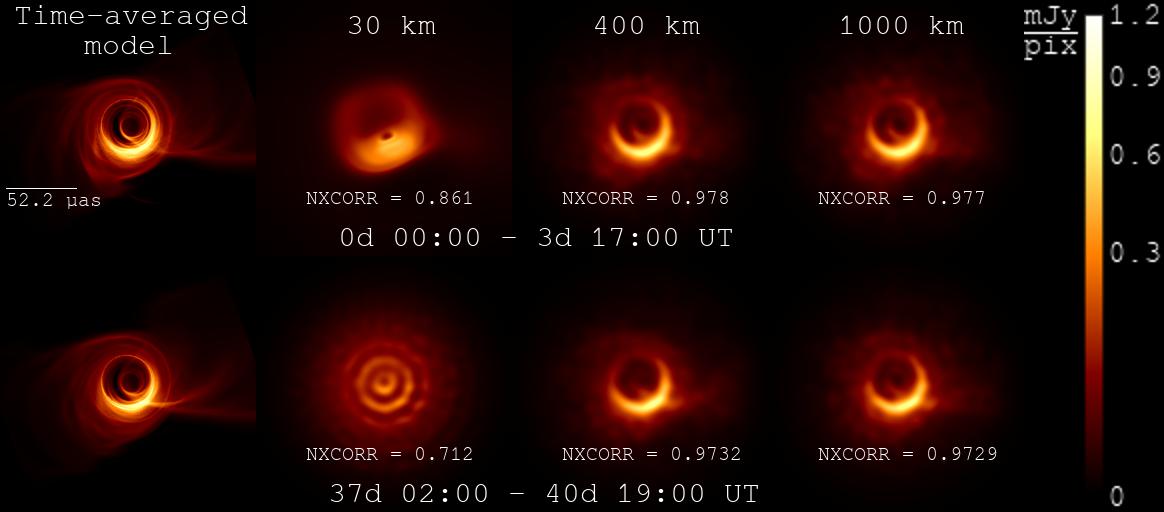}
\caption{Effect of the orbital separation on the reconstruction quality. From left to right: the time-averaged over 3.7 days theoretical emission map of M87* at 230~GHz (the model with 8.9~hours between frames); frames of the movies simulated for the EHI with 30, 400, and 1000~km orbital separations, each lasting 3.7~days, reconstructed using snapshot imaging (to highlight the effect). The source is varying during the simulated observation. Colours indicate brightness/pixel in mJy (square root scale).}
\label{fig:imgEHIsep}
\end{figure*}

\begin{figure}
\resizebox{\hsize}{!}{\includegraphics{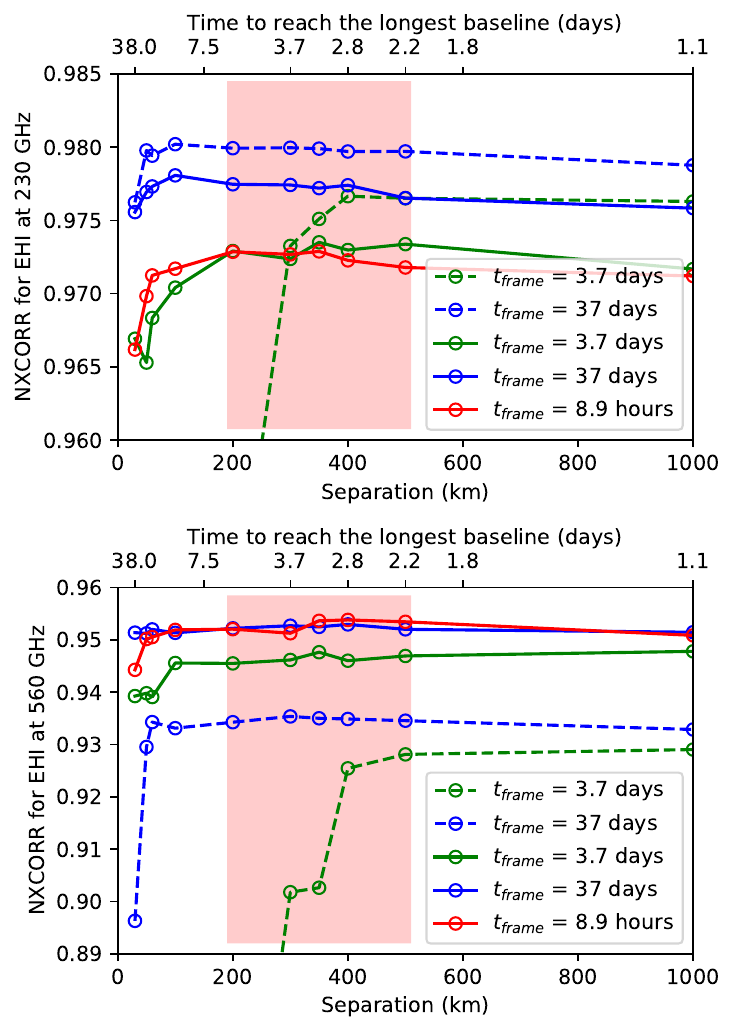}}
\caption{Quality of the M87* shadow movies obtained with different orbital separations of the EHI system setup. The movie quality is shown with the averaged normalized cross-correlation against the true image, or NXCORR, at two frequencies: (1) 230~GHz, shown in the top panel; (2) 560~GHz, shown in the bottom panel. Red, green and blue lines correspond to reconstructed movies with temporal resolutions of 8.9~hours, 3.7~days and 37~days, respectively. Dashed lines correspond to snapshot imaging; solid lines correspond to dynamical imaging. The red semitransparent area indicates orbital separations with the best quality of reconstructed movies, based on the quality of individual frames in the movies.}
\label{fig:plot230560}
\end{figure}

\begin{figure*}[!h]
\sidecaption
\includegraphics[width=12cm]{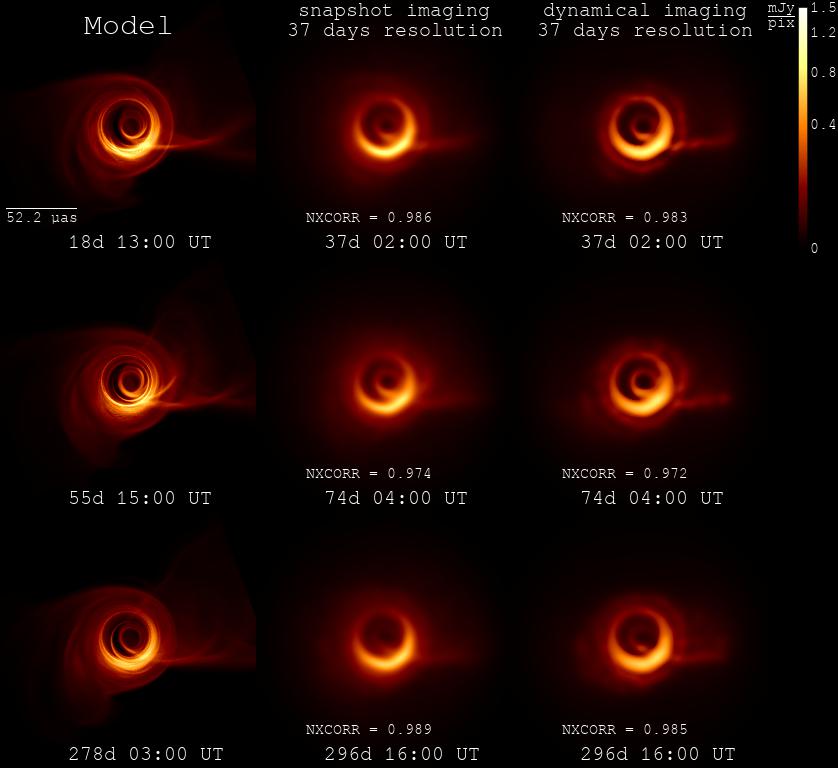}
\caption{Reconstruction of a simulated M87* shadow observation with the EHI at 230~GHz for 400~km orbital separation. From left to right: the middle theoretical emission map in the period of 37 days (the model with 3.7~days between frames); frames of the simulated movies, each lasting 37~days, reconstructed using snapshot imaging and dynamical imaging methods. For the EHI observing at 230~GHz, these movies demonstrate the best quality, according to the NXCORR, among all reconstructions that image source dynamics. The source is varying during the simulated observation. Colours indicate brightness/pixel in mJy (square root scale). The full reconstruction is available as an online movie.}
\label{fig:img230}
\end{figure*}

\begin{figure*}[!h]
\sidecaption
\includegraphics[width=12cm]{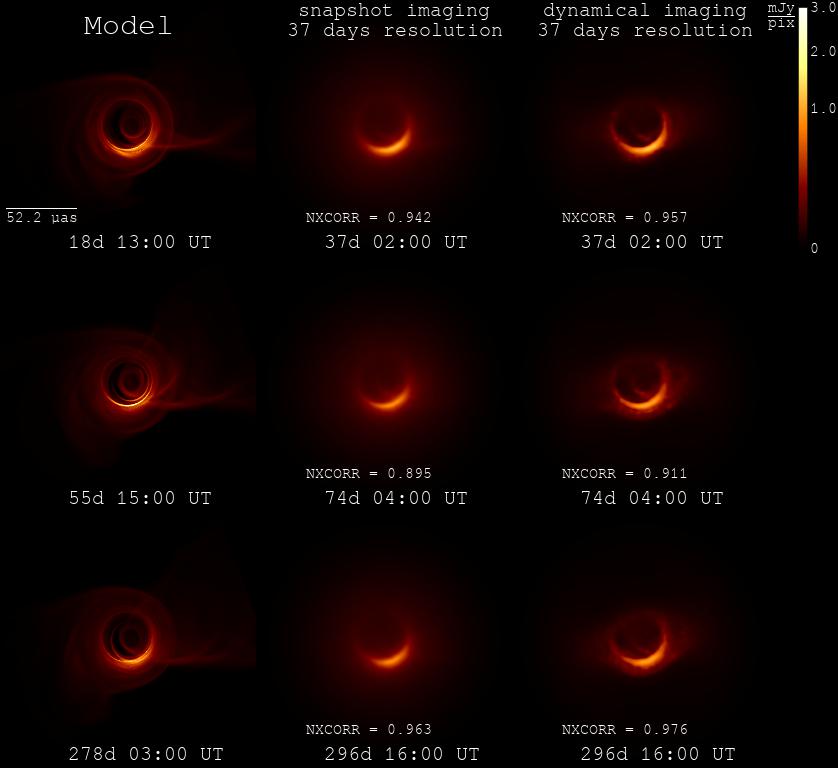}
\caption{Same as Fig.~\ref{fig:img230}, but for an observation frequency of 560~GHz. The full reconstruction is available as an online movie.}
\label{fig:img560}
\end{figure*}

\subsubsection{The EHI: 230 and 560 GHz} \label{sec:EHI}
    In Figure~\ref{fig:imgEHImethod}, we illustrate the quality of individual frames of movies reconstructed with snapshot and dynamical imaging methods. Snapshot imaging reconstructs each frame of the movie independently. Hence the absence of baselines results in an inability to reconstruct some frames with satisfactory quality, especially for the short orbital separation used. At the same time, dynamical imaging gets information for image reconstruction from other frames and, therefore, provides a visually noticeable enhancement of the image quality throughout the entire movie. Some improvement can be achieved with lower temporal resolution (Figure~\ref{fig:imgEHItscale}). The longer duration of movie frames corresponds to bigger parts of the $uv$-spiral in each snapshot. Nevertheless, apart from a stronger violation of the static source assumption, a decrease in temporal resolution implies a loss of information about the source variability. An increase of the orbital separation leads to a more uniform $uv$-coverage and a simultaneous decrease in its density (discussed in Sec.~\ref{sec:uv}). As demonstrated in Figure~\ref{fig:imgEHIsep}, the quality of frames in movies obtained with large orbital separations is significantly higher than with a small separation. In regions of high intensity, the accuracy and fidelity of movies are similar for large orbital separations. However, some distortions and artefacts corresponding to the $uv$-coverage sparsity appear at regions with low intensity, when the separation is too wide. The visually observable difference is confirmed by the image quality metrics. Since the discussed issues of the reconstruction quality depend on the uniformity and density of the $uv$-plane coverage, they are relevant for all considered frequencies. \\
    \indent Figure~\ref{fig:plot230560} shows the averaged quality of the movie reconstructions depending on the orbital separation for observations with the EHI system at 230~GHz on the top panel and 560~GHz on the bottom panel. As discussed earlier, small orbital separations provide coverage of the $uv$-plane per frame in a limited range of baselines. When the separation is wide enough to ensure a comparatively uniform $uv$-coverage per each frame, the average quality of the movie is plateauing. Outside of the best quality range of separations, the averaged quality of movies stays on a plateau due to individual frames with high image quality, while the majority of frames demonstrate slightly reduced quality, compared to movies within the best quality range. The best results for all simulations displayed in Figure~\ref{fig:plot230560} are obtained for orbital separations between 200 and 500~km, based on the quality of individual frames in the movies, although the source variability can be reconstructed reasonably well for a wider range of separations. The NXCORR metric indicates that simulations of observations at 230~GHz with 100~$t_{\mathrm{G}}$ temporal resolution provide a better quality of movies than ones with higher temporal resolutions reconstructed with the same method. It also shows a slight advantage in favour of the snapshot imaging reconstruction method compared to dynamical imaging. Since the spatial frequency is proportional to the observational frequency considering the same baselines, the $uv$-coverage is less dense at 560~GHz compared to 230~GHz with the same system setup. Therefore, dynamical imaging provides better quality of the reconstructed movies at this frequency for all tested temporal resolutions with a small difference in quality between the latter. \\
    \indent Figures~\ref{fig:img230} and \ref{fig:img560} show several frames of the movies from the simulated observation at 230~GHz and 560~GHz respectively with 400~km orbital separation and 100~$t_{\mathrm{G}}$ temporal resolution. These movies demonstrate the best quality among all reconstructions that image source dynamics at given frequencies according to the NXCORR metric. Although some temporal model brightness distribution changes are blurred out due to the temporal resolution, the characteristic shape of the features matches the model across the corresponding observational period. The right part of the images is dominated by the emission from the jet, while structures on the left part of the images display the movement of the plasma in the accretion disk. Therefore, the demonstrated detailed imaging of the changes around the shadow and at the beginning of the jet provides high-quality information about M87* dynamics. The total flux of M87* at 560~GHz is lower than at 230~GHz (Table~\ref{tab:models}) which makes the signal-to-noise ratio significantly lower and affects the quality of the images. However, the spatial resolution is higher and the environment closer to the event horizon can be probed. An improvement in the quality of movies may be reached with a different selection of reconstruction parameters, which were chosen based on simulations at 230~GHz in this work (see Appendix~\ref{sec:appendix1}). Besides, observations at 560~GHz can be improved significantly by increasing the system sensitivity (Sec.~\ref{sec:EHIplus}).

\begin{figure*}
\sidecaption
\includegraphics[width=12cm]{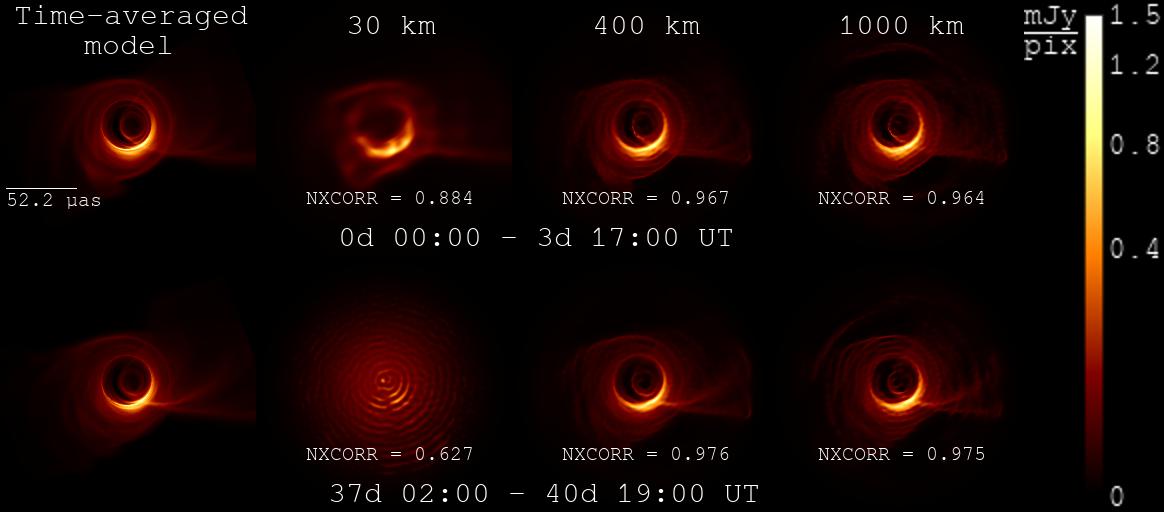}
\caption{Same as Fig.~\ref{fig:imgEHIsep}, but for observations with the EHI+ at 560~GHz.}
\label{fig:imgEHI+sep}
\end{figure*}

\begin{figure}
\resizebox{\hsize}{!}{\includegraphics{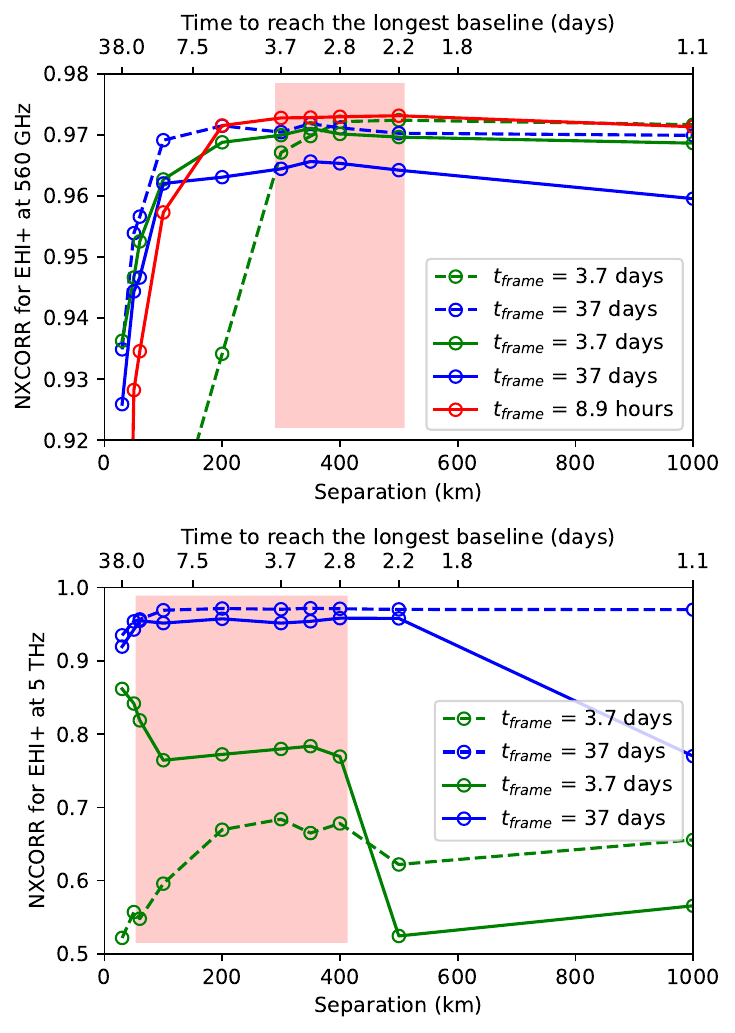}}
\caption{Same as Fig.~\ref{fig:plot230560}, but for the EHI+ system setup observing at 560~GHz (the top panel) and 5~THz (the bottom panel).}
\label{fig:plot5605}
\end{figure}

\begin{figure*}
\centering
\includegraphics[width=17cm]{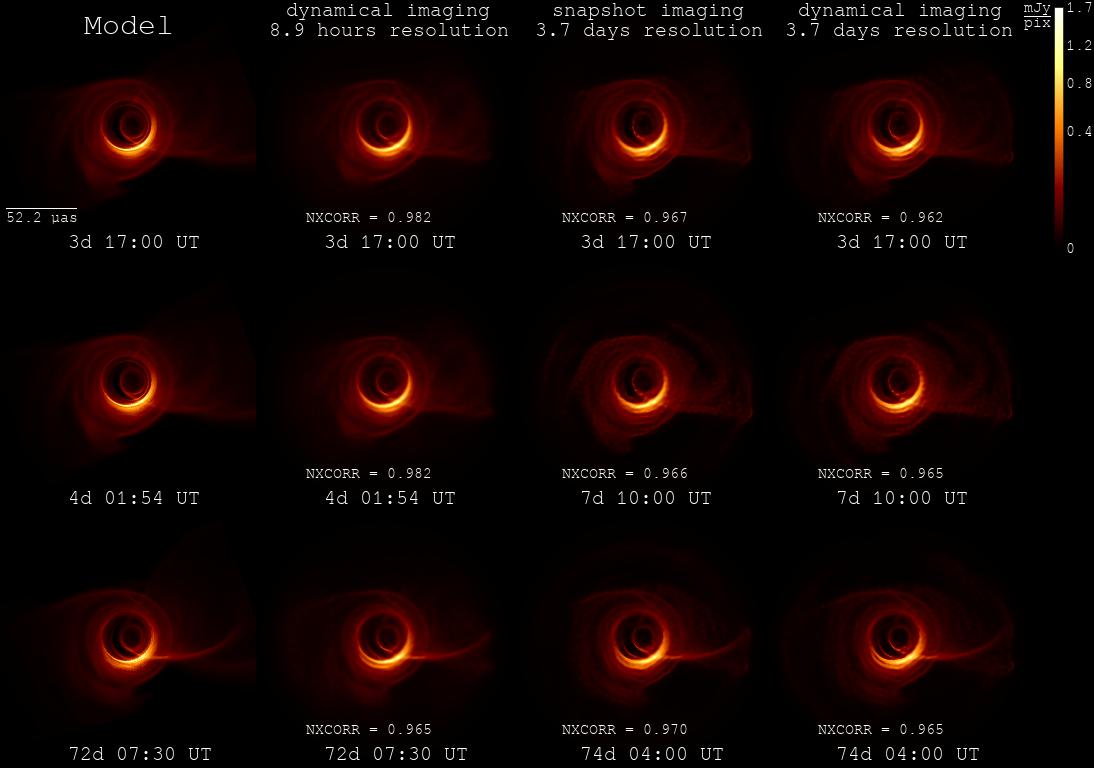}
\caption{Same as Fig.~\ref{fig:img560}, but for the EHI+. Simulated movies with frames lasting 8.9~hours are reconstructed using the dynamical imaging method; ones with frames lasting 3.7~days are reconstructed using snapshot imaging and dynamical imaging methods. The full reconstruction is available as an online movie.}
\label{fig:img560+}
\end{figure*}

\begin{figure*}
\sidecaption
\includegraphics[width=12cm]{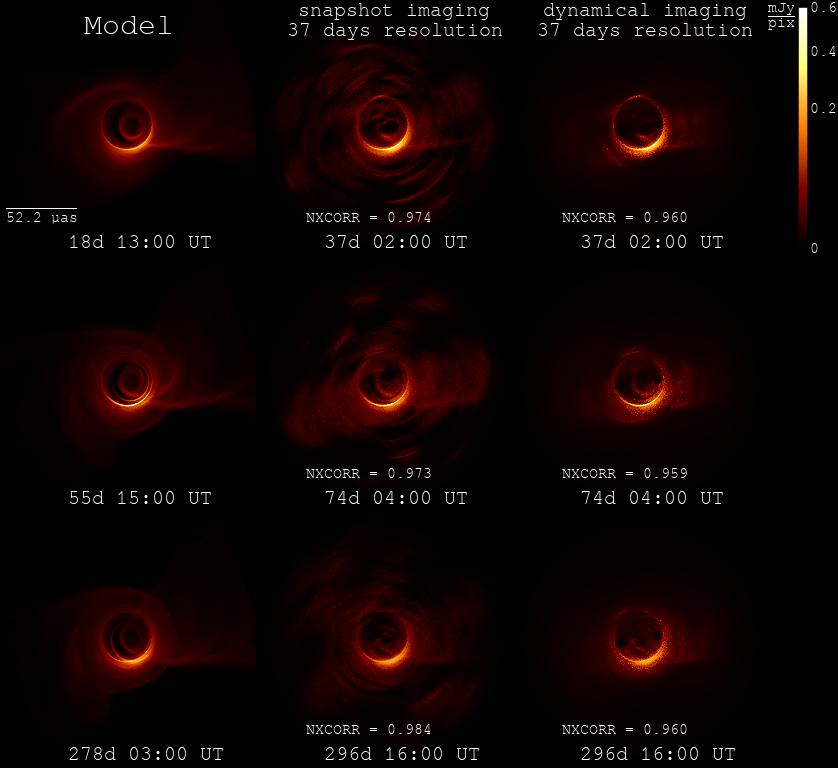}
\caption{Same as Fig.~\ref{fig:img230} and~\ref{fig:img560}, but for the EHI+ with 200~km orbital separation, observing at 5~THz. The full reconstruction is available as an online movie.}
\label{fig:img5+}
\end{figure*}

\subsubsection{The EHI+: 560 GHz and 5 THz} \label{sec:EHIplus}
    The main trends in the image quality identified for the EHI system and associated with the $uv$-coverage (Figures~\ref{fig:imgEHImethod},~\ref{fig:imgEHItscale} and \ref{fig:imgEHIsep}) remain relevant for the EHI+ system. Moreover, Figure~\ref{fig:imgEHI+sep} shows that reconstruction artefacts and distortions observed at the largest orbital separations become clearer with an increasing observation frequency due to a general decrease in $uv$-spiral density. At the same time, the effect of the lower flux density at higher frequencies is compensated by the lower noise in the EHI+ system. \\
    \indent Figure~\ref{fig:plot5605} shows the quality of the image reconstruction depending on the orbital separation for observations with the EHI+ system at 560~GHz on the top panel and 5~THz on the bottom panel. All simulations of observations with the EHI+ at 560~GHz have a better quality of reconstructed movies compared to observations with the EHI since EHI+ data have a higher signal-to-noise ratio. Moreover, all displayed simulations demonstrate a similar quality regardless of the reconstruction method and the temporal resolution. This shows that, despite the issue of the $uv$-coverage per frame discussed above, the system provides coverage that is dense and uniform enough for the very sharp and accurate source reconstruction. The best results for all displayed simulations of observations with the EHI+ at 560~GHz are obtained for orbital separations between 300 and 500~km, based on the quality of individual frames in the movies. Further increase of the observation frequency to 5~THz exacerbates the $uv$-coverage issue due to the extra reduction of its density. Additional imaging difficulties are introduced by the onward total flux decrease (see Table~\ref{tab:models}). As demonstrated in Figure~\ref{fig:plot5605}, data received in 10~$t_{\mathrm{G}}$ (3.7~days) becomes insufficient to ensure satisfactory quality of movies for the source variability imaging. The temporal resolution of 100~$t_{\mathrm{G}}$ (37~days) improves image quality to some extent. In the case of observations at 5~THz, the best quality range of separations lies between 60 and 400~km. The quality of movies obtained with orbital separations out of this range is insufficient for the detailed imaging of the changes in the M87* environment due to the unstable quality of frames in movies. Nevertheless, a 5~THz observation frequency can provide a resolution of $\sim0.5$~$\mu$as, which significantly enhances the imaging capabilities of the system. \\
    \indent Figure~\ref{fig:img560+} shows several frames of the movies from the simulated observation at 560~GHz with 400~km orbital separation and temporal resolutions of 1~$t_{\mathrm{G}}$ and 10~$t_{\mathrm{G}}$. Although variability is hard to notice even in the model on 1~$t_{\mathrm{G}}$ timescale, the reconstructed movie reproduces features of the source with high accuracy. For 10~$t_{\mathrm{G}}$ temporal resolution, the snapshot imaging reconstruction method gives slight artefacts, which mostly disappear with the dynamical imaging method. The temporal resolution of 100~$t_{\mathrm{G}}$ provides a lower quality of the movies since reconstruction artefacts intensify. Observations with 200~km orbital separation and 100~$t_{\mathrm{G}}$ temporal resolution reached the best quality among all at 5~THz. As seen in Figure~\ref{fig:img5+}, both methods are able to reconstruct the main features of the source in regions of high intensity. In low-intensity regions, movies demonstrate artefacts, however, dynamical imaging significantly reduces them. Additional improvement of the reconstruction quality may be reached with the dynamical imaging regularizer $\mathcal{R}_{flow}$, which assumes that an image evolves according to a steady flow of the flux density over time \citep{Johnson2017}. In summary, all the reconstructed movies demonstrate that the EHI concept allows for observations of the M87* dynamics from space at frequencies up to 5~THz with the signal-to-noise ratio as the main limiting parameter. Moreover, the angular resolution at 5~THz can provide exceptionally precise measurements of black hole parameters and remarkably deep tests of different theories of gravity, as well as the additional increase in the number of sources with resolvable shadows.

\begin{figure}
\resizebox{\hsize}{!}{\includegraphics{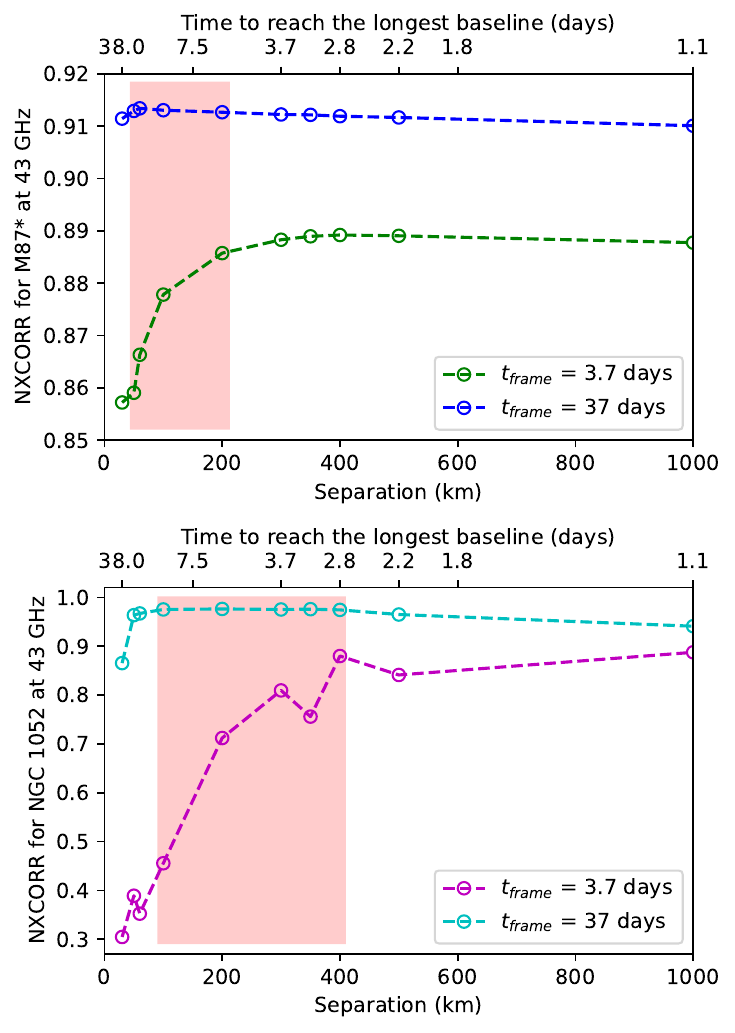}}
\caption{Quality of black hole jet movies obtained with different orbital separations of the EHI system setup at 43~GHz. The movie quality is shown with the averaged NXCORR, for two sources: (1) M87*, shown in the top panel; (2) NGC\,1052, shown in the bottom panel. Green and magenta lines correspond to reconstructed movies with a 3.7-day temporal resolution; blue and cyan lines correspond to reconstructed movies with a 37-day temporal resolution. The red semitransparent area indicates orbital separations with the best quality of reconstructed movies, based on the quality of individual frames in the movies.}
\label{fig:plot43}
\end{figure}

\begin{figure}
\resizebox{\hsize}{!}{\includegraphics{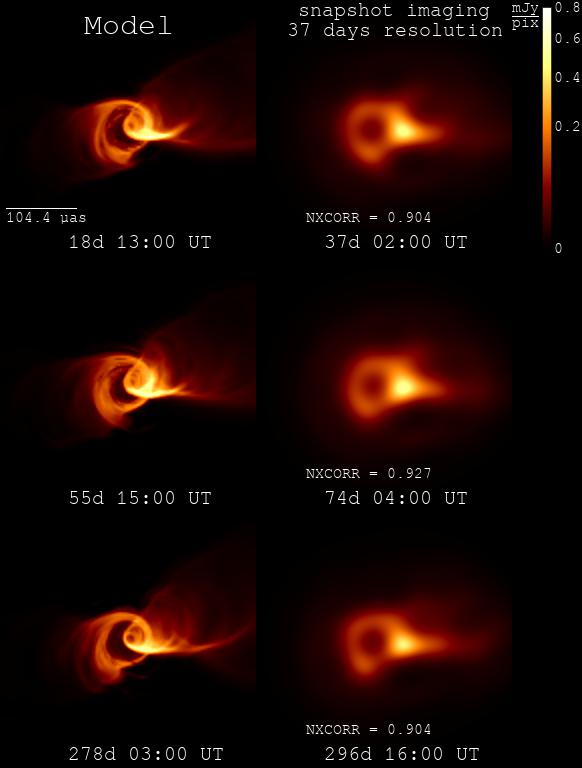}}
\caption{Reconstruction of a simulated M87* jet observation with the EHI at 43~GHz for 60~km orbital separation. From left to right: the middle theoretical emission map in the period of 37~days (the model with 3.7~days between frames); frames of the simulated movie, each lasting 37~days, reconstructed using the snapshot imaging method. For the EHI observing at 43~GHz, this movie demonstrates the best quality, according to the NXCORR, among all reconstructions that image source dynamics. The source is varying during the simulated observation. Colours indicate brightness/pixel in mJy (square root scale). The full reconstruction is available as an online movie.}
\label{fig:img43m87}
\end{figure}

\begin{figure}
\resizebox{\hsize}{!}{\includegraphics{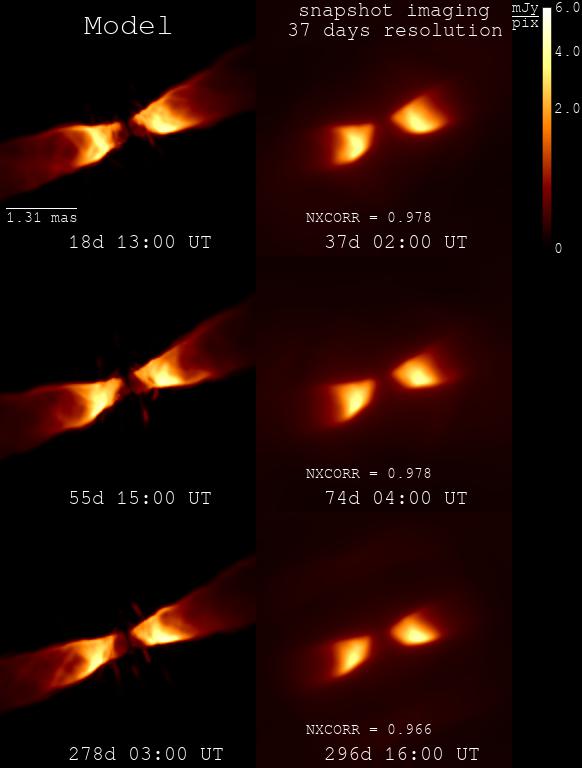}}
\caption{Same as Fig.~\ref{fig:img43m87}, but for NGC\,1052, observed with 200~km orbital separation. The full reconstruction is available as an online movie.}
\label{fig:img43ngc1052}
\end{figure}

\subsection{Imaging of an AGN jet} \label{sec:jet}
    The possible secondary EHI science goal is imaging the variability in the extended jets of various AGNs at 43~GHz. In this subsection, we research the EHI imaging capability and the optimal orbital separation for this goal on the example of M87* and NGC\,1052. Simulations of imaging the M87* jet at 43~GHz were produced similarly to the simulations of the M87* shadow imaging at 230 and 560~GHz. Observations of models with 1 and 10~$t_{\mathrm{G}}$ time intervals between frames were reconstructed with temporal resolutions of 10 and 100~$t_{\mathrm{G}}$, respectively, considering the variability of the source during the observation. For coarse models of NGC\,1052 (see Sec.~\ref{sec:maps}), simulations were produced analogously. Observations of models with 8.9 and 89~hours between frames were reconstructed with temporal resolutions of 3.7 and 37~days per frame, respectively. All simulations were reconstructed using the snapshot imaging method. \\
    \indent The image quality of reconstructions becomes more sensitive to the $uv$-coverage issue (see Sec.~\ref{sec:shadow}) for the imaging of extended jets. Figure~\ref{fig:plot43} shows the quality of reconstructed movies depending on the orbital separation for M87* on the top panel and NGC\,1052 on the bottom panel. As seen from M87* and NGC\,1052 jet imaging simulations, a relatively high temporal resolution (3.7~days in this project) is insufficient for adequate observations. The $uv$-coverage obtained during this period lacks the required amount of baselines, and only a few frames in a movie can be reconstructed with satisfactory quality. At the same time, movies with approximately a month-long (37~days) temporal resolution demonstrate significantly better quality. In the case of M87*, the best quality of reconstructed movies corresponds to orbital separations in a range from 50 to 200~km, based on the quality of individual frames in the movies. For NGC\,1052, high-quality movies are produced with orbital separations between 100 and 400~km. Orbital separations out of this range provide abundant reconstruction artefacts. \\
    \indent Figure~\ref{fig:img43m87} shows several frames of the movie reconstructed from simulated observations of M87* with 60~km orbital separation and 100~$t_{\mathrm{G}}$ temporal resolution. The spatial resolution at 43~GHz frequency is significantly lower, but changes in the brightness distribution at the beginning of the extended jet are visible and can be used to reconstruct jet dynamics. The noticeable deficit of the observed intensity in the central region of images can be confused with the shadow but it corresponds to the spine of the jet directed away from the observer. There are limitations in the modelling of extended jets since actual sources demonstrate much more extended emission compared to our model. Thus, the signal-to-noise ratio is reduced in our simulations, which decreases the quality of reconstructed movies. The best result for simulated observation of the NGC\,1052 jet is reached with 200~km orbital separation. Figure~\ref{fig:img43ngc1052} shows several frames of the corresponding movie with clear changes in the jet shape and the brightness distribution along it from frame to frame. The changes are most noticeable in the brightest parts of the jet. For a more detailed study, physically correctly scaled models are required in order to consider the timescales of the source variability.

\begin{figure*}
\sidecaption
\includegraphics[width=12cm]{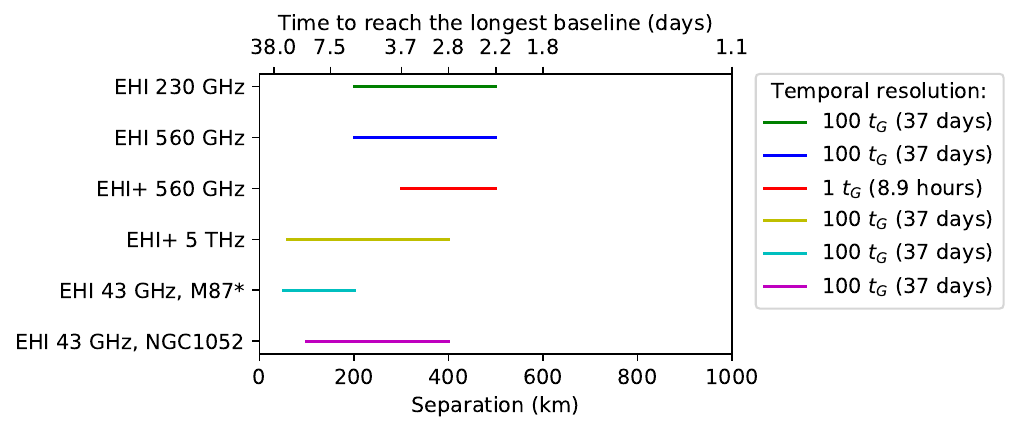}
\caption{Orbital separations and temporal resolutions providing the best quality imaging of the source. The EHI system consists of three 4.0-metre antennas with standard system parameters; the EHI+ system consists of three 15.0-metre antennas with more optimistic noise parameters. Simulated observations at 230, 560~GHz, and 5~THz image the M87* shadow; simulated observations at 43~GHz image M87* and NGC\,1052 jets.}
\label{fig:allthebest}
\end{figure*}

\section{Conclusions and summary} \label{sec:conclusions}
    In this paper, we have presented simulations of the EHI SVLBI system consisting of three satellites in circular MEOs with slightly different radii \citep{Martin-Neira2017, Kudriashov2019, Roelofs2019, Kudriashov2021b}. The proposed accurate relative positioning of the satellites and interchange of local oscillator signals allow for the use of complex visibilities. The absence of atmospheric data corruptions allows for imaging at frequencies significantly higher compared to ground-based observations. In this work, we investigated the effects of different orbital separations on the reconstructed image quality. The orbital separation can be changed during the mission to the optimal separation for the currently observed source and frequency. \\
    \indent The EHI setup provides a spiral-shaped sampling of the $uv$-plane and can perform high-fidelity imaging of the M87* shadow without additional changes in the satellite motion during observations. Nevertheless, dividing the $uv$-coverage into snapshots, considered in this paper, requires a balance between density and homogeneity in each snapshot that influences the selection of the orbital separation. Too small separations provide coverage of the $uv$-plane per frame in an overly narrow range of baselines and too large ones produce excessively sparse $uv$-coverage, which results in distortions and artefacts in the reconstructed movies. An increase in the observation frequency also enlarges the $uv$-coverage sparsity. \\
    \indent Using GRMHD simulations of M87* and model system parameters, similar to those discussed in \citet{Roelofs2019}, we performed simulated long-lasting observations to assess the quality of movies that can be expected. For the imaging of the structural variations of the M87* environment, an important potential EHI science goal, simulations were performed with temporal resolutions of 1, 10, and 100~$t_{\mathrm{G}}$ (8.9~hours, 3.7~days, and 37~days) at 230 and 560~GHz for the standard EHI system, 560~GHz and 5~THz for the EHI+ system with more optimistic noise parameters. Also, simulations for the EHI were performed with temporal resolutions of 3.7 and 37 days at 43~GHz for another potential goal, imaging the structural variations of extended jets. The simulated reconstructed movies with temporal resolutions of 3.7 and 37~days included source variability during each frame of the resulting movies. \\
    \indent Figure~\ref{fig:allthebest} summarizes results of our simulations. The M87* shadow high-fidelity imaging is possible at 230 and 560~GHz with the orbital separation in the range from 200 to 500~km for the EHI. The best quality of the reconstructions has been obtained by observations with 400~km orbital separation and 100~$t_{\mathrm{G}}$ (37~days) temporal resolution at both frequencies. Since the resolvable variability of the source has approximately a monthly timescale, a 37-day temporal resolution is optimal for observations with the standard EHI system. Simulations of the M87* shadow imaging at 560~GHz with the EHI+ system demonstrate the highest accuracy and fidelity with the orbital separation in the range from about 300 to 500~km and the best performance for 400~km separation. The signal-to-noise ratio of observations with the EHI+ system at 560~GHz allows for the M87* environment variability imaging with 1~$t_{\mathrm{G}}$ (8.9~hours) temporal resolution. Changes in the M87* environment are hardly resolvable on the gravitational timescale in the model itself, nevertheless, a decrease in the temporal resolution is not favourable since it produces artefacts in the reconstructed movies. In the case of observations with the EHI+ at 5~THz, simulations only with orbital separations in the range from 60 to 400~km and 100~$t_{\mathrm{G}}$ (37~days) temporal resolution demonstrate the sufficient quality of movies for the detailed imaging of the changes in the M87* environment. The best result at 5~THz, obtained with 200~km orbital separation, shows accurate imaging of the source main features in high-intensity regions, however, the $uv$-coverage sparsity at such a high frequency leads to strong reconstruction artefacts in low-intensity regions. Nevertheless, the angular resolution at 5~THz allows for extraordinarily precise measurements. \\
    \indent The imaged changes in the brightness distribution around the shadow and at the beginning of the jet are expected to provide information sufficient to reconstruct the dynamics of the source. Such deep probes of the surrounding environment of M87* can allow for deeper tests for general relativity and alternative theories of gravity since these theories make predictions of the appearance of the radio emission generated by material falling into the black hole. Moreover, more accurate models of a black hole environment can be tested. The emission at a higher frequency originates closer to the event horizon, however, the M87* total flux is gradually decreasing at frequencies higher than 230~GHz \citep{Davelaar2019}. Simulated observations demonstrate that the reduction of the system noise level noticeably improves the quality of reconstructed movies and makes source changes accessible for imaging on shorter timescales and at higher frequencies. Therefore, the signal-to-noise ratio is the main parameter limiting the EHI in observations of the M87* shadow, which needs to be considered to optimize the efficiency and the design of the system in general. \\
    \indent The resolution at 43~GHz is not high enough to resolve the shadow of M87* but sufficient for imaging changes of the brightness distribution at the beginning of the M87* jet. Orbital separations in the range from 50 to 200~km and 100~$t_{\mathrm{G}}$ (37~days) temporal resolution are expected to be the most favourable for observations with the best result for 60~km separation. The EHI capability for imaging jets of other sources was performed using NGC\,1052 as a generic example. The M87* models were scaled to match the field of view and the total flux. This scaling is not physical because the black hole mass and hence the variability timescales were not properly scaled when changing the angular size of the source, therefore, additional research is essential. Nevertheless, we have demonstrated that jet observations are exacting to orbital separation and require lower temporal resolution to obtain sufficient $uv$-coverage. Observations with a separation in the range from 100 to 400~km and a 37-day temporal resolution can capture changes in the jet shape and the brightness distribution along it. The best result has been achieved with 200~km orbital separation. Therefore, EHI observations at 43~GHz can provide sufficient information to reconstruct structural variations of the relativistic jets of M87* and other AGNs, for example, NGC\,1052, 3C 273, 3C 84, M81* or Centaurus~A. These observations can improve our understanding of the jet launching and collimation processes. \\
    \indent Apart from M87*, the EHI is capable of imaging Sgr\,A*, as discussed by \citet{Roelofs2019}. Moreover, shadows of some other sources can potentially be resolved by the EHI, such as the shadows of supermassive black holes at the centres of the Sobrero Galaxy M104 and the elliptical galaxy M84 \citep{Johannsen2012}. The EHI+ system observing at 5~THz could potentially resolve shadows of M81* \citep{Johannsen2012} and Centaurus~A \citep{Janssen2021}. With all these capabilities, the EHI concept will be of great astrophysical interest. \\
    \indent The implementation of the concept into an actual mission implies several technical challenges. The maximum possible orbit reconstruction accuracy depends on laser ranging, accelerometers, orbital modelling, fringe fitting, and other measures that are currently under investigation. Sending reduced data to the ground requires the elaboration of the on-board correlation and processing. Consequently, realistic observations may hinder image quality due to uncertainties not considered in this paper. Nevertheless, we demonstrate the general processes in the temporal structure of the EHI coverage. Moreover, the signal-to-noise ratio is the primary parameter limiting EHI imaging capabilities as discussed above, therefore, the possibilities of reducing the system noise should also be investigated. \\
    \indent Technical system requirements for the EHI can be relaxed with the application of closure phases to accompany visibility amplitudes instead of complex visibilities. However, closure phase calculation requires data from all three pairs of satellites for each frame of the reconstructed movies. Depending on the orbital separation, partial sampling of the $uv$-plane, considered in this paper, includes information from three pairs of satellites on different fractions of the frames on the investigated observational period. $35\%$ of appropriate frames for the 50~km orbital separation increase to $45\%$ for separations of 100 and 200~km and $75\%$ and $90\%$ for 300 and 400~km ones, respectively. Therefore, the dependence of the possibility to calculate closure phases on the orbital separations should be considered in the design of the system. \\
    \indent Additional consideration should also be given to the effect of the angle between the orbital plane and the line of sight on the imaging quality. Moreover, instead of a three-satellite system, two satellites can be considered, as it is also one of the possible configurations. Another possibility is investigating a space-space-ground hybrid system that would provide fast baseline coverage for dynamical imaging of rapidly varying sources such as Sgr\,A* \citep{Palumbo2019}.

\begin{acknowledgements}
    This work is supported by the ERC Synergy Grant ``BlackHoleCam: Imaging the Event Horizon of Black Holes'' (Grant 610058). The authors thank Manuel Martin-Neira, Volodymyr Kudriashov and Daniel Palumbo for their helpful comments and discussions on this work. We are grateful to the anonymous referee for useful and constructive comments. AS personally acknowledges Olesya Kuchay for her unlimited support during the period of writing the manuscript and Jeremy Tregloan-Reed for his invaluable advice. FR was supported by NSF grants AST-1935980 and AST-2034306. This work was supported by the Black Hole Initiative, which is funded by grants from the John Templeton Foundation (Grant \#62286) and the Gordon and Betty Moore Foundation (Grant GBMF-8273) - although the opinions expressed in this work are those of the author(s) and do not necessarily reflect the views of these Foundations. JD is supported by NASA grant NNX17AL82G and a Joint Columbia/Flatiron Postdoctoral Fellowship. Research at the Flatiron Institute is supported by the Simons Foundation. The GRMHD simulations were performed on the Dutch National Supercomputing cluster Cartesius and are funded by the NWO computing grant 16431.
\end{acknowledgements}

\bibliographystyle{aa} 
\bibliography{bibliography}

\begin{appendix} 
\section{Parameters of the image reconstruction} \label{sec:appendix1}
\subsection{Snapshot imaging} \label{sec:snapshot}
\begin{figure}
\resizebox{\hsize}{!}{\includegraphics{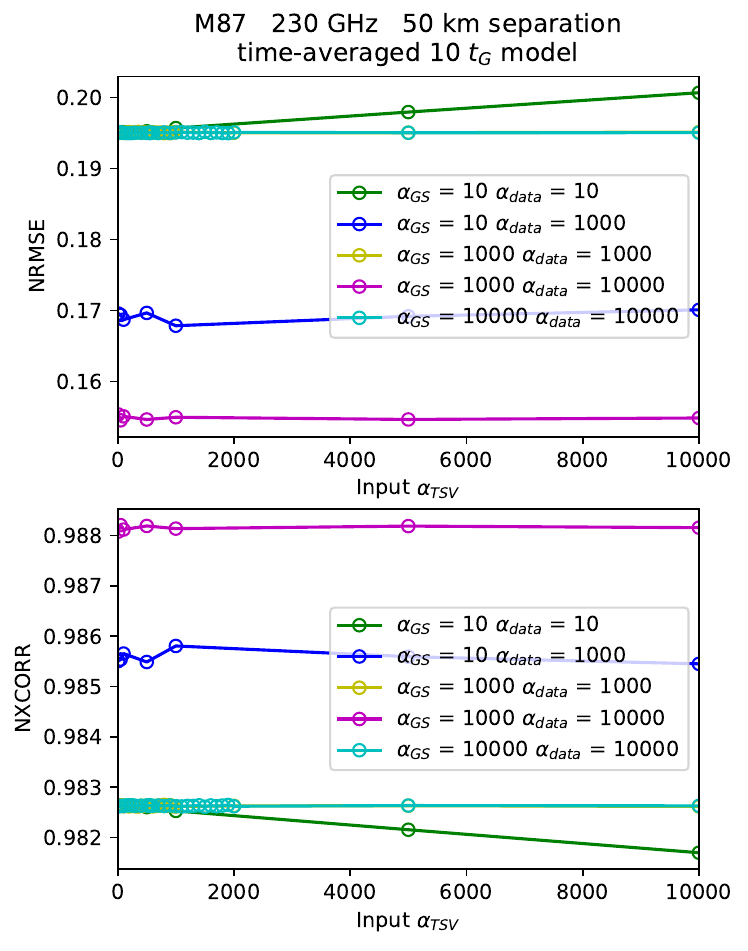}}
\caption{Quality of the reconstructions performed using the snapshot imaging method with different weighting parameters. The image quality is measured in two ways: (1) the normalized root-mean-square error against the true image, or NRMSE, shown in the top panel; (2) the normalized cross-correlation against the true image, or NXCORR, shown in the bottom panel. $\alpha_{data}$ is the data term weighting parameter, $\alpha_{GS}$ is the Gull-Skilling entropy function weighting parameter, $\alpha_{TSV}$ is the Total Squared Variation regularization weighting parameter.}
\label{fig:snapparam1}
\end{figure}

\begin{figure*}
\centering
\includegraphics[width=17cm]{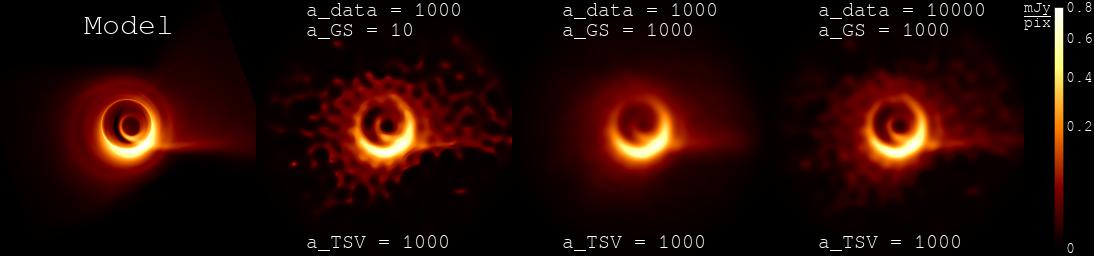}
\caption{Reconstructions, performed using the snapshot imaging method with different weighting parameters. Time-averaged M87* observation simulated at 230~GHz, 50~km orbital separation. $a\_data$ is the data term weighting parameter $\alpha_{data}$, $a\_GS$ is the Gull-Skilling entropy function weighting parameter $\alpha_{GS}$, $a\_TSV$ is the Total Squared Variation regularization weighting parameter $\alpha_{TSV}$.}
\label{fig:snapparam2}
\end{figure*}

\begin{figure}
\resizebox{\hsize}{!}{\includegraphics{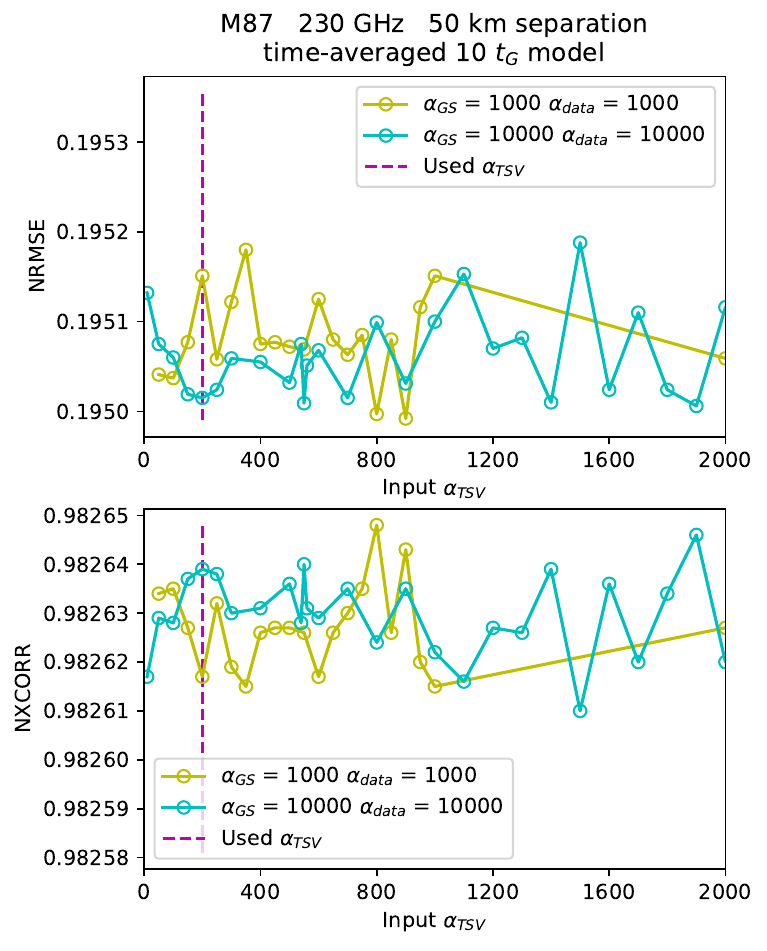}}
\caption{Same as Fig.~\ref{fig:snapparam1}, but with enlarged scale.}
\label{fig:snapparam3}
\end{figure}

    An RML algorithm minimizes the weighted sum of $\chi^2$ and a regularizer \citep{Gull1978}. $\chi^2$ is the goodness-of-fit test statistic that compares the visibilities of the test image to the data. The regularizer contains prior information of the image, such as image smoothness or sparsity. The RML algorithm, by its definition, gives an image that has no more structures than required to fit the data. For snapshot imaging, we chose to use two regularization terms. \\
    \indent One of the possible regularizers implemented into the \verb|eht-imaging| software is the Gull-Skilling entropy function \citep{GullSkilling1991}. The Gull-Skilling entropy function is a general form of the entropy function derived from a Bayesian interpretation of the RML. In combination with the Total Squared Variation regularizer, it shows the best visible result of the image reconstruction during the regularizer selection and, therefore, was picked in this work. The Total Squared Variation regularization \citep{Kuramochi2018} is the denoising process. The Total Squared Variation regularization is the modification of the Total Variation regularization \citep{RudinOsherFatemi1992}. This modification favours images with smaller variations between adjacent pixels, which leads to the smoothing of edgelike features. Apart from better consistency with observational data compared to post-processing smoothing, edge-smoothed images demonstrate a better representation of the model images in the case of diffuse astronomical objects. \\
    \indent The selection of the weighting parameters for the data term and both regularization terms was performed by simulating an observation of the time-averaged 10~$t_{\mathrm{G}}$ M87* model. The used system setup had a 50~km orbital separation and an observing frequency of 230~GHz. Firstly, the weighting parameters were varied from 10 to $10^4$. Figure~\ref{fig:snapparam1} shows the best-performing combinations of weights according to our image quality metrics. The NRMSE performs a pixel-to-pixel comparison, while the NXCORR compares the bulks of intensity patterns. Nevertheless, both metrics show similar dependencies. The best performance according to image quality metrics is achieved when the data term weighting $\alpha_{data}$ is bigger than the weighting for Gull-Skilling entropy $\alpha_{GS}$. However, reconstructed images are noticeably better when these two terms are equally weighted (Figure~\ref{fig:snapparam2}). The weighting parameter for the Total Squared Variation regularization $\alpha_{TSV}$ demonstrates little effect on the quality of reconstruction. When it is significantly less than the other two terms, the metric performance is better. However, when this condition is reached, further alteration of the Total Squared Variation regularization weight provides only a vanishingly small variation in image quality metrics (Figure~\ref{fig:snapparam3}). Thus, the weighting parameters for the data term $\alpha_{data}$ and the Gull-Skilling entropy function term $\alpha_{GS}$ were both selected to equal 1000, the weighting parameter for the Total Squared Variation regularization term $\alpha_{TSV}$ was selected to equal 200. To facilitate comparisons across different frequencies and orbital separations, the weighting parameters for the data and regularization terms are identical for all snapshot imaging reconstructions displayed in this paper.\\
    \indent For both snapshot and dynamical imaging, the image reconstruction was performed in several (typically 3-7) rounds to help the optimizer converge to a single image structure. In the first round, a circular Gaussian with a FWHM of 55~$\mu$as (700~$\mu$as for NGC\,1052 jet simulations) and total flux equal to that of the corresponding model frame were used as the initial and prior image. In the following imaging rounds, the result of the previous round was blurred to the nominal array resolution and used as a new initial and prior image. \\

\begin{figure}
\resizebox{\hsize}{!}{\includegraphics{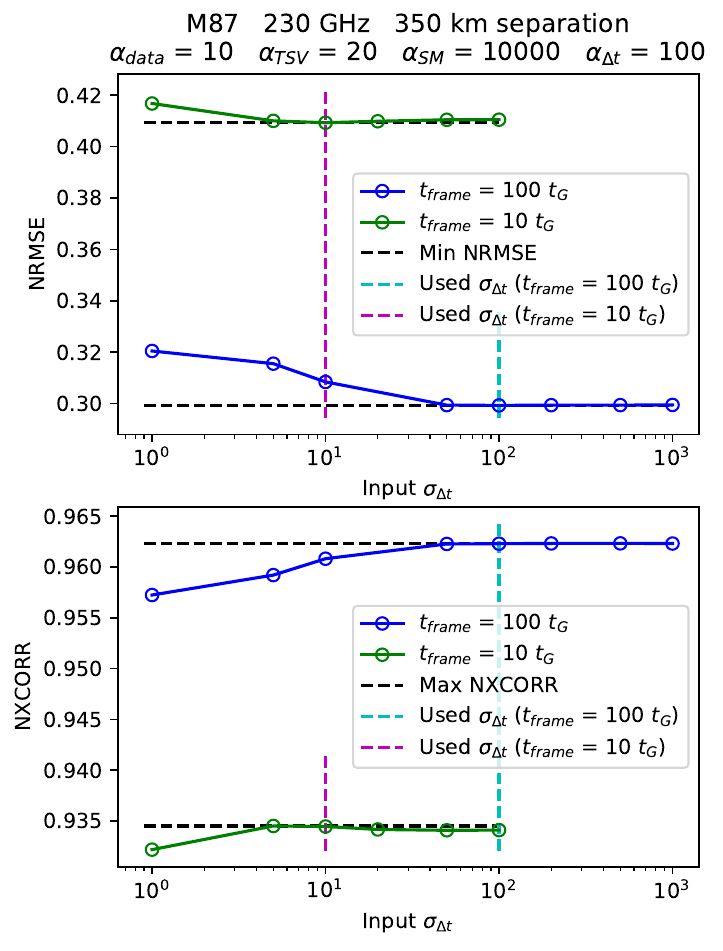}}
\caption{Quality of the M87* movies reconstructed using the dynamical imaging method with different values of the dynamical regularizer parameter $\sigma_{\Delta t}$. The image quality is measured in two ways: (1) the NRMSE, shown in the top panel; (2) the NXCORR, shown in the bottom panel. Green and blue lines correspond to reconstructed movies with temporal resolutions of 3.7 days and 37 days, respectively. $\alpha_{data}$ is the data term weighting parameter, $\alpha_{TSV}$ is the Total Squared Variation regularization weighting parameter, $\alpha_{SM}$ is the Second Moment regularization weighting parameter, $\alpha_{\Delta t}$ is the dynamical regularizer $\mathcal{R}_{\Delta t}$ weighting parameter.}
\label{fig:dynparam4}
\end{figure}

\begin{figure}
\resizebox{\hsize}{!}{\includegraphics{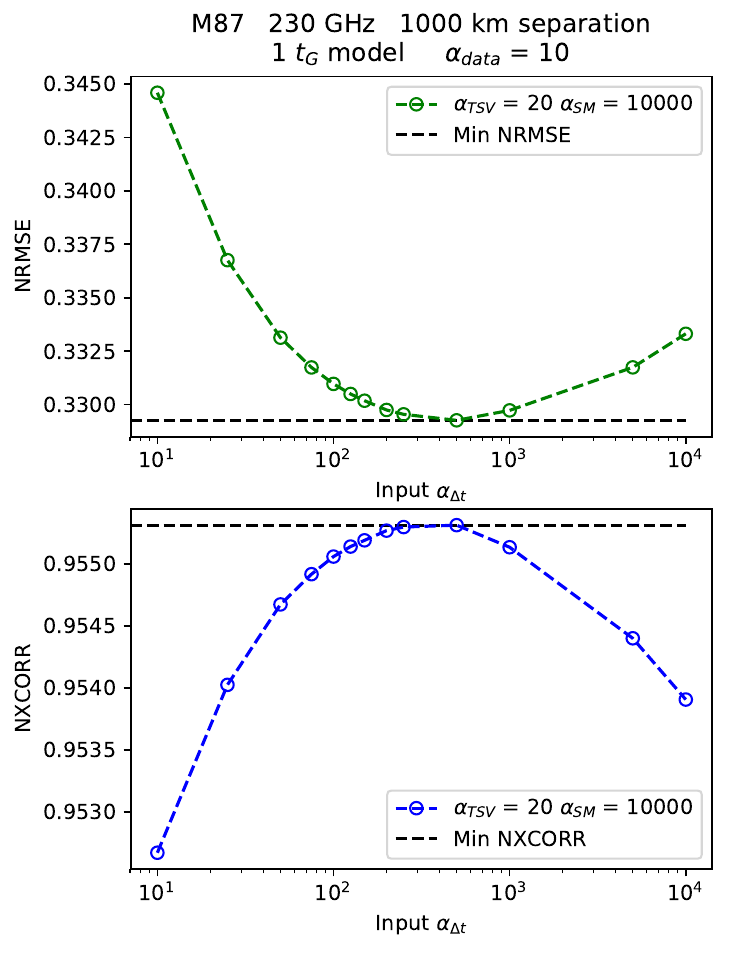}}
\caption{Quality of the M87* movies reconstructed using the dynamical imaging method with different values of the weighting parameter of the dynamical regularizer $\mathcal{R}_{\Delta t}$. The image quality is measured in two ways: (1) the NRMSE, shown in the top panel; (2) the NXCORR, shown in the bottom panel. $\alpha_{data}$ is the data term weighting parameter, $\alpha_{SM}$ is the Second Moment regularization weighting parameter, $\alpha_{TSV}$ is the Total Squared Variation regularization weighting parameter.}
\label{fig:dynparam3}
\end{figure}

\begin{figure}
\resizebox{\hsize}{!}{\includegraphics{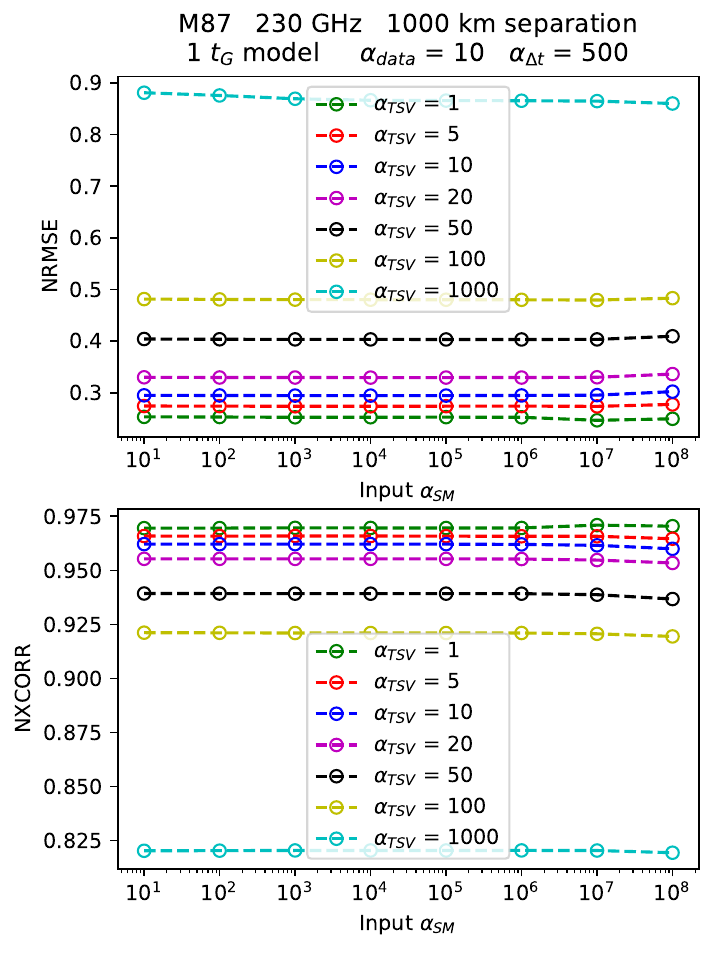}}
\caption{Quality of the M87* movies reconstructed using the dynamical imaging method with different weighting parameters of static regularizers. The image quality is measured in two ways: (1) the NRMSE, shown in the top panel; (2) the NXCORR, shown in the bottom panel. $\alpha_{data}$ is the data term weighting parameter, $\alpha_{\Delta t}$ is the dynamical regularizer $\mathcal{R}_{\Delta t}$ weighting parameter, $\alpha_{SM}$ is the Second Moment regularization weighting parameter, $\alpha_{TSV}$ is the Total Squared Variation regularization weighting parameter.}
\label{fig:dynparam1}
\end{figure}

\begin{figure}
\resizebox{\hsize}{!}{\includegraphics{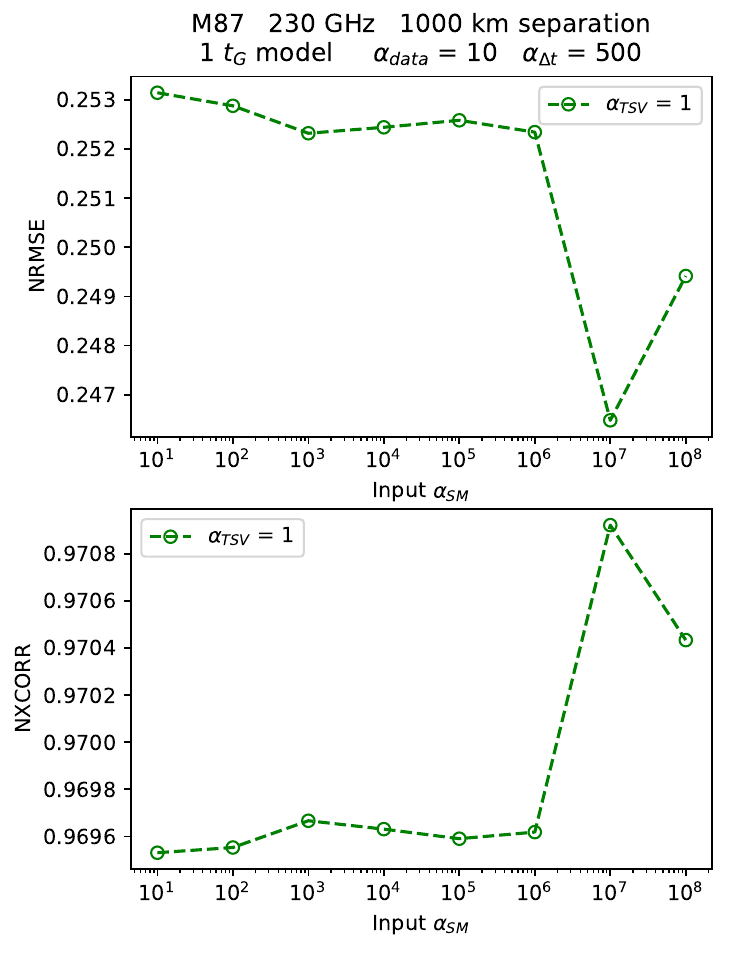}}
\caption{Same as Fig.~\ref{fig:dynparam1}, but with enlarged scale.}
\label{fig:dynparam2}
\end{figure}

\subsection{Dynamical imaging} \label{sec:dynamical}
    Dynamical imaging assumes that the images in the set are connected and gets information for image reconstruction from other frames \citep{Johnson2017}. This allows the missing data to be partially inherited from the rest of the movie by interpolation. In this work, a combination of the dynamical regularization with two static regularization terms was used. \\
    \indent A generic regularizer $\mathcal{R}_{\Delta t}$ from \citet{Johnson2017}, used in this paper, assumes that the images in the set are temporally connected, each being a small perturbation of the previous frame. This enforces the continuity of features from frame to frame in the reconstructed movie. For the static regularization terms, we chose the Second Moment regularizer \citep{Issaoun2019} and the Total Squared Variation regularizer (described in Appendix~\ref{sec:snapshot}). The Second Moment regularization constrains the spread of flux density in reconstructed images to a motivated region defined by the user. In the case of the EHI system, although single frames can have very limited coverage, the dense $uv$-coverage is produced in the whole range of baselines considering time-averaged data. Hence source parameters, required for the Second Moment regularization, can be obtained from the same observational data by the time-averaged image reconstruction. The implementation of the Second Moment regularizer into the dynamical imaging method can grant the highest completeness, accuracy and fidelity of reconstructed movies. \\
    \indent The dynamical regularizer $\mathcal{R}_{\Delta t}$ computes the similarity between frames of the movie by calculating the summed difference of the reconstructed flux density of pixels among all adjacent frames after blurring using a circular Gaussian with standard deviation $\sigma_{\Delta t}$ \citep{Johnson2017}. The selection of the parameter $\sigma_{\Delta t}$ was performed for each temporal resolution independently. This parameter equals 0 for observations with 1~$t_{\mathrm{G}}$ temporal resolution. The limit $\sigma_{\Delta t} \to 0$ is appropriate when the expected motion between consecutive frames is smaller than the finest resolution of reconstructed features comparable to the nominal array spatial resolution \citep{Johnson2017}. To choose $\sigma_{\Delta t}$ for observations with 10~$t_{\mathrm{G}}$ temporal resolution, half the number of frames of the M87* model with 10~$t_{\mathrm{G}}$ interval between frames was used. For observations with 100~$t_{\mathrm{G}}$ temporal resolution, the full model with 10~$t_{\mathrm{G}}$ interval between frames was used considering the variability of the source during the observation. The used system setup had 350~km orbital separation and an observing frequency of 230~GHz. $\sigma_{\Delta t}$ was varied from 1 to 100 and from 1 to 1000 for reconstructions with averaging over 10 and 100~$t_{\mathrm{G}}$ per frame, respectively. According to image quality metrics averaged over the duration of reconstructed movies, $\sigma_{\Delta t}$ was selected to equal 10 and 100, respectively, in these two cases. (Figure~\ref{fig:dynparam4}). \\
    \indent The selection of weighting parameters for the dynamical imaging method was performed by simulating observation of half the number of frames of the M87* model with 1~$t_{\mathrm{G}}$ interval between frames. The used system setup had 1000~km orbital separation for the widest representation of baselines on each of the frames, the observing frequency was 230~GHz. The data term weighting $\alpha_{data}$ was left default and equals 10. The weighting parameter for dynamical regularization $\alpha_{\Delta t}$ was ranged from 10 to $10^4$. As $\alpha_{\Delta t} \to 0$ the dynamical imaging becomes equivalent to snapshot imaging with corresponding static regularizers. Taking $\alpha_{\Delta t} \to \infty$ leads to the reconstruction of the time-averaged image \citep{Johnson2017}. According to averaged over the duration of reconstructed movies image quality metrics, $\alpha_{\Delta t}$ was selected to equal 500 (Figure~\ref{fig:dynparam3}). The weighting parameters for the Second Moment regularizer $\alpha_{SM}$ and for the Total Squared Variation regularizer $\alpha_{TSV}$ were ranged between 10 and $10^8$, and between 1 and $10^3$, respectively. Figure~\ref{fig:dynparam1} demonstrates image quality metrics averaged over the duration of reconstructed movies for different combinations of these parameters. Dependencies demonstrated by the NRMSE and the NXCORR are identical, similar to the snapshot imaging method. The best reconstruction quality corresponds to the smallest weight of the Total Squared Variation regularization. The weighting of the Second Moment regularization produces little effect on the quality of reconstruction with a slight improvement for large values, as shown in Figure~\ref{fig:dynparam2}. Therefore, weighting parameters for the Second Moment regularization $\alpha_{SM}$ and the Total Squared Variation regularization $\alpha_{TSV}$ were selected to equal $10^7$ and 1, respectively. The regularization parameters are the same for all dynamical imaging reconstructions displayed in this paper.
\end{appendix}
\end{document}